\documentclass[a4paper,
              ]{jacow}
\usepackage{mathtools}
\makeatletter%                           % test for XeTeX where the sequence is by default eps-> pdf, jpg, png, pdf, ...
\ifboolexpr{bool{xetex}}                 % and the JACoW template provides JACpic2v3.eps and JACpic2v3.jpg which might generates errors
 {\renewcommand{\Gin@extensions}{.pdf,%
                    .png,.jpg,.bmp,.pict,.tif,.psd,.mac,.sga,.tga,.gif,%
                    .eps,.ps,%
                    }}{}
\makeatother
\usepackage{epstopdf}
\ifboolexpr{bool{xetex} or bool{luatex}} % test for XeTeX/LuaTeX
 {}                                      % input encoding is utf8 by default
 {\usepackage[utf8]{inputenc}}           % switch to utf8

\usepackage[USenglish]{babel}

\ifboolexpr{bool{jacowbiblatex}}%        % if BibLaTeX is used
 {%
  \addbibresource{jacow-test.bib}
  \addbibresource{biblatex-examples.bib}
 }{}

\begin{document}
%\linenumbers
\title{Study of Phase Reconstruction Techniques applied to Smith-Purcell Radiation Measurements}
%\thanks{Work supported by the French ANR (contract ANR-12-JS05-0003-01), the PICS (CNRS) "Development of the instrumentation for accelerator experiments, beam monitoring and other applications and  Research Grant \#F58/380-2013 (project F58/04) from the State Fund for Fundamental Researches of Ukraine in the frame of the State key laboratory of high energy physics." }

\author{
N. Delerue~\thanks{delerue@lal.in2p3.fr}, J. Barros,   LAL, Orsay, France PARIS SACLAY\\
O. Bezshyyko, V. Khodnevych, Taras Shevchenko National University of Kyiv, Ukraine}

\maketitle

\begin{abstract}
Measurements of coherent radiation at accelerators typically give the absolute value of the beam profile Fourier transform but not its phase. Phase reconstruction techniques such as Hilbert transform or Kramers Kronig reconstruction are used to recover such phase. We report a study of the performances of these methods and how to optimize the reconstructed profiles. 
\end{abstract}

\section{Longitudinal bunch profile measurement at particle accelerators}

On a particle accelerator the longitudinal profiles of a particle bunch can not easily be measured. Several indirect measurement techniques have been established relying on the measurement of the spectrum of radiation emitted by the bunch either when it crosses a different material~\cite{OTR_LURE} or when it passes near a different material~\cite{ODR_Cianchi,Doucas_ESB}. This emitted  spectrum encode the longitudinal profile through the relation:

\begin{equation}
I(\lambda) = I_1(\lambda) ( N + | F(\lambda) |^2 N^2 )
\end{equation}

where $I(\lambda)$ is the emitted intensity as a function of the wavelength $\lambda$.  $I_1(\lambda)$ is the intensity of the signal emitted by a single particle and $F(\lambda)$ is a form factor that encodes the longitudinal and transverse shape of the particle bunch. Recovering the longitudinal profile requires to invert this equation however this is not straightforward as the information about the phase of the form factor can not be measured and therefore is not available.

A phase reconstruction algorithm must therefore be used to recover this phase. Several methods exist (see for example~\cite{KK}). 
In this article we describe how we implemented two of these methods and compared their performances.

\section{Reconstruction methods}

When it is only possible to measure the amplitude of the complex signal, it is necessary to recover the phase of the available data. 
We assume that the function of the longitudinal beam density is analytical. 
For an analytic function this is easier because the real and imaginary part are not completely independent.
The Kramers-Kronig relations~\cite{KK} helps restore the imaginary part of an analytic function $\varepsilon(\omega)$ from its real part and vice versa.

% In this case, the relations are as follows:
%$$\varepsilon_1(\omega) = {1 \over \pi} \mathcal{P}\!\!\!\int \limits_{-\infty}^\infty {\varepsilon_2(\omega') \over \omega' - \omega}\,d\omega'$$
%and
%$$\varepsilon_2(\omega) = -{1 \over \pi} \mathcal{P}\!\!\!\int \limits_{-\infty}^\infty {\varepsilon_1(\omega') \over \omega' - \omega}\,d\omega',$$
% where $\varepsilon(\omega) = \varepsilon_1(\omega) + i \varepsilon_2(\omega)$  and $\mathcal{P}$ denotes the Cauchy principal value. 

To recover the phase from the amplitude, the function should  be written as: $log(F(\omega))=log(\rho(\omega))+i\Theta(\omega)$ with $\rho(\omega)$ its amplitude and $\Theta(\omega)$ its phase. 
The Kramers-Kronig relations can then be applied as follows:
\begin{equation}
\Theta(\omega_0)  =  \frac{2\omega_0}{\pi} \textit{P}\int^{+ \infty}_{0}\frac{ln(\rho(\omega) )}{\omega_0^2-\omega^2}d\omega
\end{equation}
The basis of this relationship are the Cauchy-Riemann conditions (analyticity of function).  %In this case, the value spectrum can gain value at [0, $\infty$).\par

In some cases this phase can also be obtained simply by using the Hilbert transform of the spectrum:
\begin{equation}
\Theta(\omega_0)  =  -\frac{1}{\pi} \textit{P}\int^{+ \infty}_{- \infty}\frac{ln(\rho(\omega))}{\omega_0-\omega}d\omega.
\end{equation}
{As the Hilbert transform ($\textit{H}$) is related to the Fourier transform (${\cal F}$): 
\begin{equation}
{\cal F}(\textit{H}(u))(\omega)=(-isgn(\omega)){\cal F}(u)(\omega),
\end{equation}
the calculation of phase can use an optimised FFT code and is therefore much faster than calculating the Kramers-Kronig's integral.}
We implemented in Matlab these two different phase reconstruction methods. The Hilbert transform method has the advantage of being directly implemented in Matlab, allowing a much faster computing.

\section{Description of the simulations}

To test the performance of these methods we created a small Monte-Carlo program that randomly simulates profiles (${\cal G} (x)$) made of the combination of 5 gaussians according to the formula $ {\cal G} (x)= \sum_{i=1}^{5}  A_i  \exp{\frac{-(\frac{x}{mX} - \mu_i)^2 }{2 \sigma^2_i}} $ where $mX=2^{16}$ and $A_i$, $\mu_i$ and $\sigma_i$ are random numbers with  $x \in [1;mX]$, $A_i \in [0;1] $, $\mu_i \in 0.5 + [ -11.44 ; +11.44  ] \times 10^{-9} $ and  $\sigma_i \in [3;9] \times 10^{-9}$ . {The values of these ranges have been chosen to generate profiles that are not disconnected (that is profiles whose intensity drops to almost zero between two peaks) without being perfect gaussian.
% So the value of $\mu_i$ must be in the same order and approximately equal  $\sigma_i$(if distance between to peaks is less than $\sigma_i$, it can not be separated, if greater -- a lot of disjoint peaks can happen). 
We checked that our conclusions are valid across this range. }
% VH add and change text

Using this formula we generated 1000 profiles, then took the absolute value of their Fourier transform $ {\cal F} = \| \mbox{FFT} \left( {\cal G}\right) \|$ and  sampled at a limited number of frequency points ($F_i = {\cal F}(\omega_i)$) as would be done with a real experiment in which the number of measurement points is limited (limited number of detectors or limited number of scanning steps).

To estimate the performance of the reconstruction several estimators are available. We choose to use the $\chi^2$, defined as follow:
\begin{equation}
\chi^2=\sum_i\omega_i^2(O_i-E_i)^2/N,
\end{equation}
where $O_i$ is the observed value , $E_i$ is the expected (simulated) value, $\omega_i=1/\sqrt{O_i+E_i}$ is the weight of the point, N is the number of points.\par
\begin{figure}[htbp]
 \centering
  \includegraphics*[width=70mm]{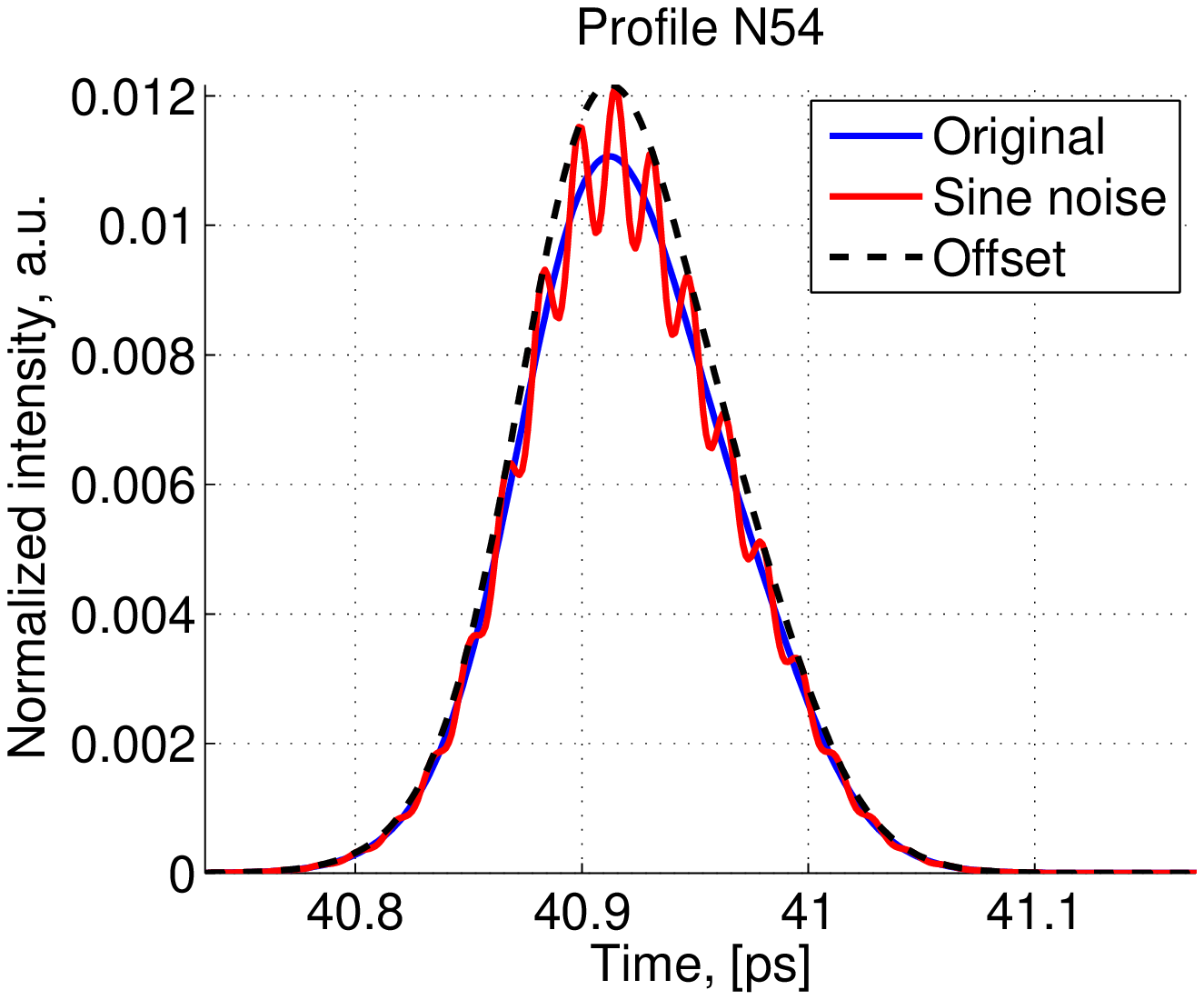}
  \caption{Example of profiles giving very different $\chi^2$ despite being relatively similar. \\
  $\chi^2_{\mbox{sine noise}}=3.8219\times 10^{-8}, \chi^2_{\mbox{offset}}=7.2661\times 10^{-8}$; For profile with sine noise: FW0.1M=0.0241, FW0.2M=0.044 FWHM=0.0621 FW0.8M=0.1849 FW0.9M=0.3619. As FWXM calculated from top of profile, for all profiles  FWXM=0. }% VH quetion 5
   \label{Offsine}
\end{figure}

However two very similar profiles but with a slight offset, will give a worse $\chi^2$ than a profile with oscillations (see figure \ref{Offsine}). This can be partly mitigated (in the case of horizontal offset) by offsetting one profile with respect to the other until the $\chi^2$ is minimized.

Also we decided to look at the FWHM which was generalized as FWXM where $X \in [0.1 ; 0.9]$ is the fraction of the maximum value at which  the full width of the reconstructed profile was calculated (with this definition the standard FWHM is noted FW0.5M).  We created an estimator $\Delta_{FWXM}$ defined as follow:

%$$
%\Delta_{FWXM} = \mbox{Max}_{X \in \mbox{rset} }\left| \frac{FWXM_{\mbox{orig}} - FWXM_{\mbox{reco}}}{FWXM_{\mbox{orig}} }\right| 
%$$
\begin{equation}
\Delta_{FWXM} = \left| \frac{FWXM_{\mbox{orig}} - FWXM_{\mbox{reco}}}{FWXM_{\mbox{orig}} }\right|_{X \in  \{ 0.1 ; 0.2 ; 0.5 ; 0.8 ; 0.9\} } 
%\Delta_{FWXM} = \sum_{X \in  \{ 0.1 ; 0.2 ; 0.5 ; 0.8 ; 0.9\} } \left| \frac{FWXM_{\mbox{orig}} - FWXM_{\mbox{reco}}}{FWXM_{\mbox{orig}} }\right| 
\end{equation}
where 
%$\mbox{rset} = \{ 0.1 ; 0.2 ; 0.5 ; 0.8 ; 0.9\}$, 
$FWXM_{\mbox{orig}}$ and $FWXM_{\mbox{reco}}$ are the FWXM of the original and reconstructed profiles respectively.

Here two profiles that are similar but slightly offset (in position or amplitude) will nevertheless return good values of this estimator despite returning a rather large $\chi^2$.

\begin{figure}[htbp]
 \centering
  \includegraphics*[width=70mm]{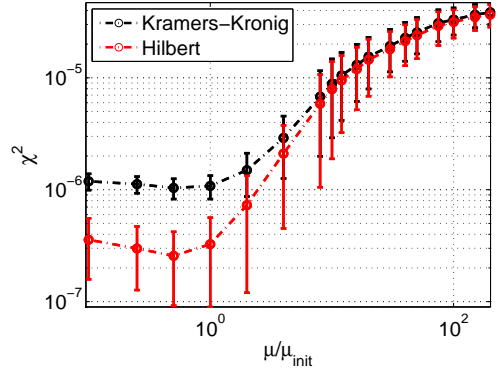}\\
  \includegraphics*[width=70mm]{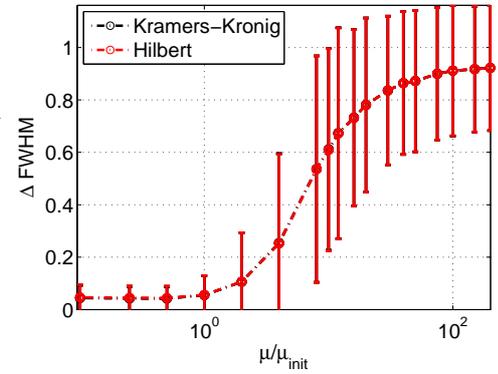}\\
  \includegraphics*[width=70mm]{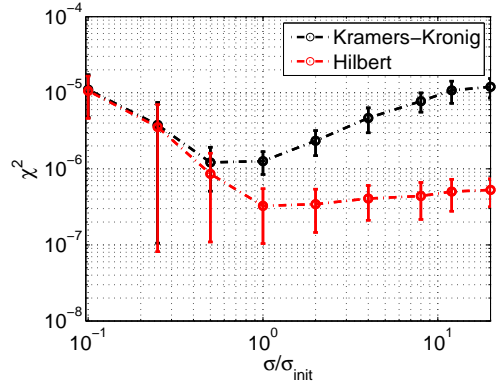}\\
  \includegraphics*[width=70mm]{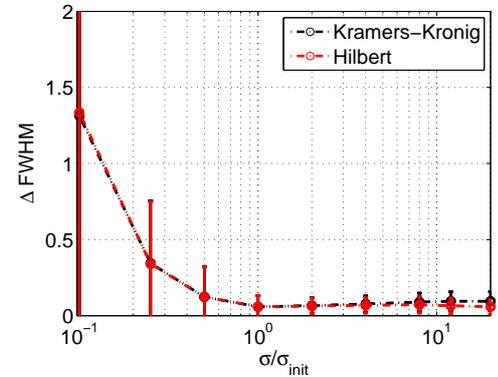}\\
  \caption{The constraints on the parameters $\sigma_i$ (top) and $\mu_i$ (bottom) due to effect of scaling in terms of the $\chi^2$ and $Delta$FWHM ratio. For each point 1000 simulations were made.}%VH question 8
   \label{sigma_chi2}
\end{figure}

To ensure that the choice of the parameters $\sigma_i$ and $\mu_i$ for the simulations does not bias significantly the results, their value has been varied and this is shown in figure~\ref{sigma_chi2}.

\begin{figure}[htbp]
 \centering
  \includegraphics*[width=70mm]{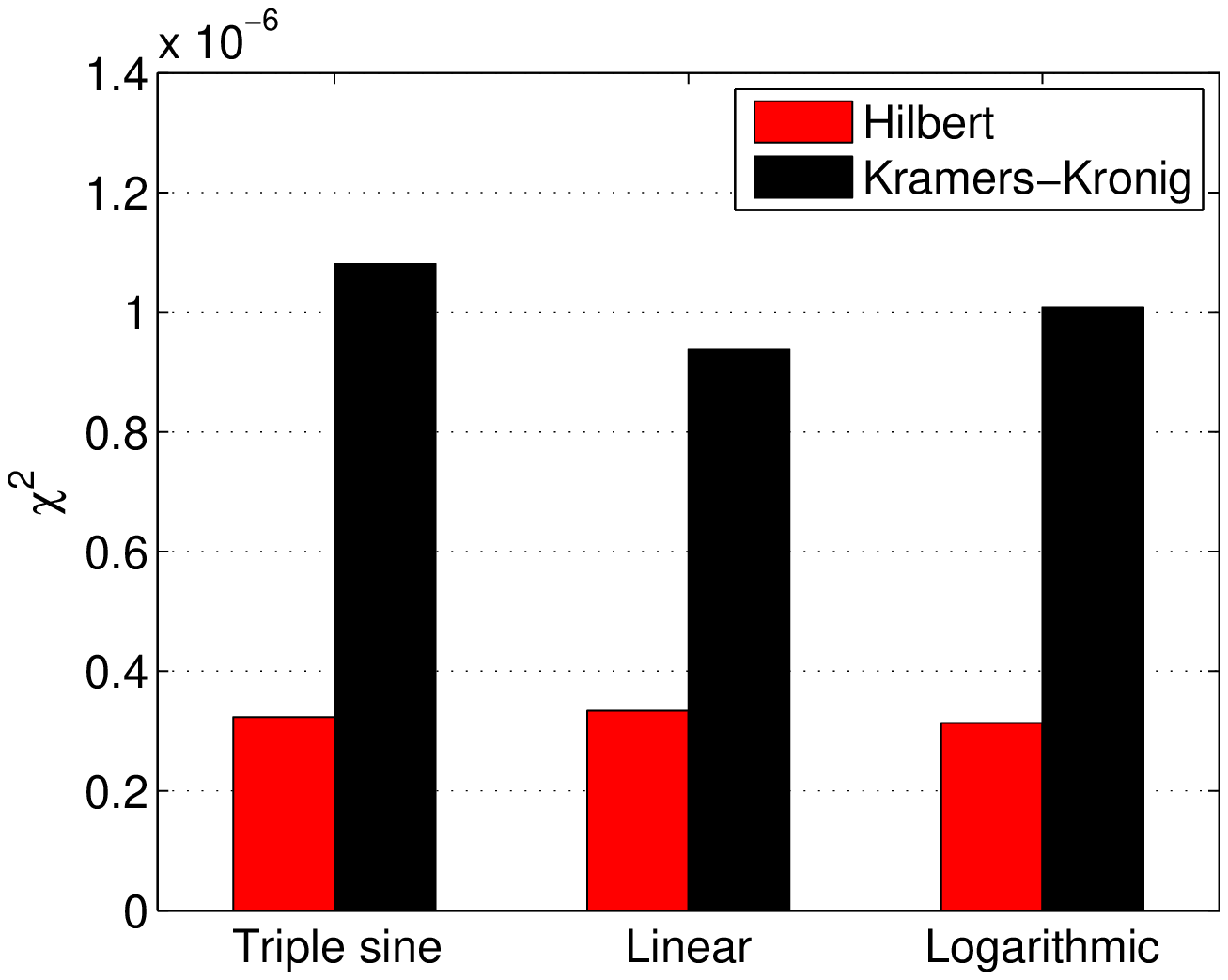} \\
    \includegraphics*[width=70mm]{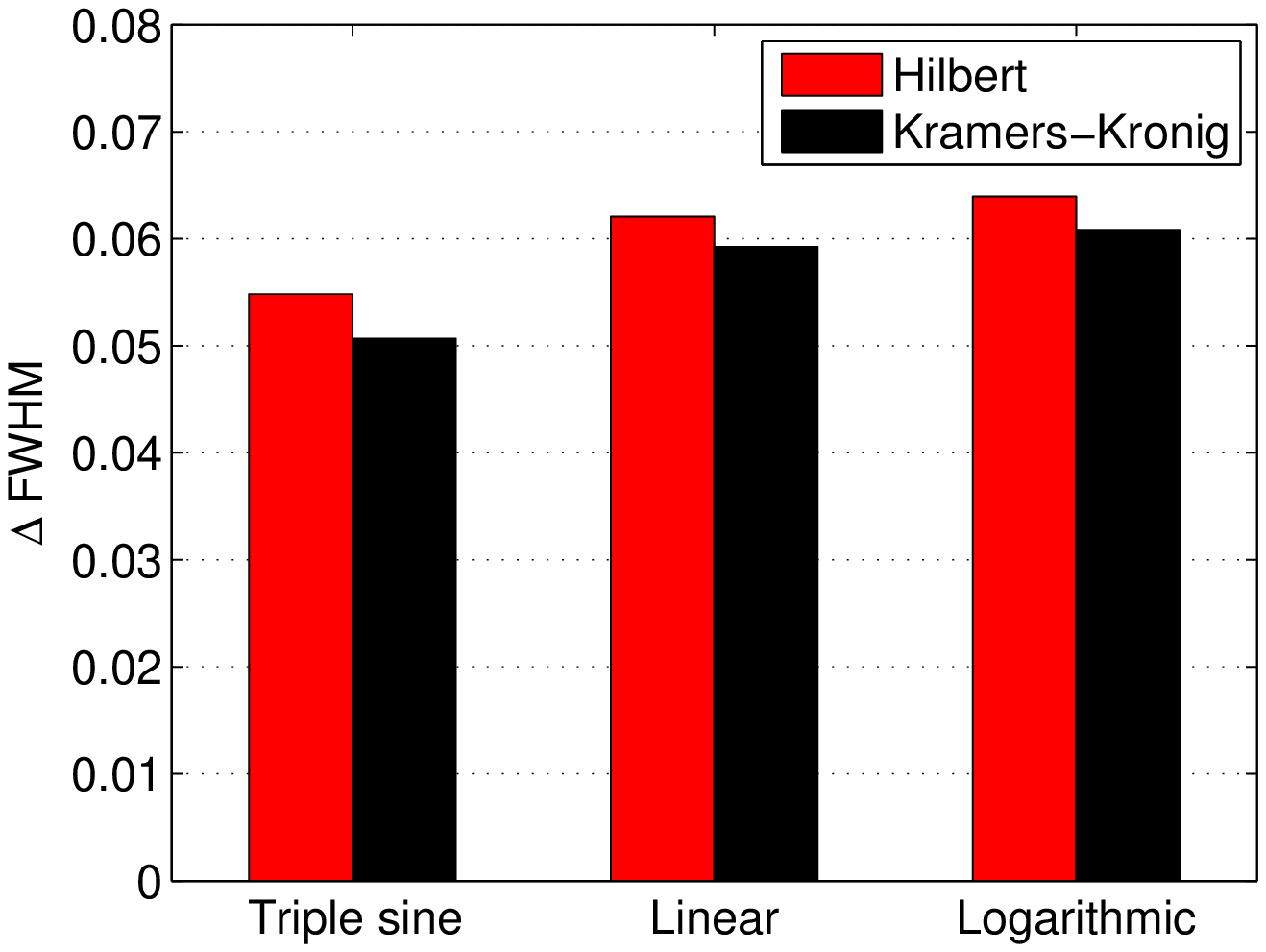}
  \caption{Comparison of different samplings with $\chi^2$ criterium (top) and $\Delta_{FWHM}$ (bottom).}
   \label{samp}
\end{figure}

Different distributions have been used for the frequencies $\omega_i$: linear, logarithmic, triple-sine. 
In most sampling schemes we used 33 frequencies  to make it comparable with the Triple-sine distribution used in~\cite{E203prstab}. These sampling schemes are defined as follow:
\begin{itemize}
\item \textit{Triple-sine} This sampling matches that of the E-203 experiment at FACET~\cite{E203prstab}. Eleven detectors are located every $10^o$ around the interaction point and 3 different sets of wavelengths are used, giving the following distribution:
\begin{equation} \label{eq:lamb}
\frac{c}{\omega_i}=l_n (1-cos(\Theta_i )) 
\end{equation}
with $l_n =50, 250, 1500 \mu m$ and $\Theta_i$ varying between $40^o$ and $140^o$ by steps of $10^o$.
%Uniform location of detectors in space corresponds to the inhomogeneous sampling frequency and vice versa. We cho
%So next sampling is linear in frequecy.
\item \textit{Linear sampling} Here sampling points are distributed uniformly. The first and last points of sampling are the first ($\omega_0$) and last ($\omega_f$) points used in the Triple-sine sampling. The following formula gives the sampling frequencies:
\begin{equation}
\omega_i=\omega_0+(\omega_f-\omega_0)/32\times[0:1:32].
\end{equation}
\item \textit{Logarithmic sampling}. Here sampling points are distributed  logarithmically:
\begin{equation}
\omega_0*exp(log(\omega_f/\omega_0)\times[0:1:32]/32).
\end{equation}
For this sampling also the first and last points are the same as in Triple-sine sampling. 
\end{itemize}
{The study of the sampling is important, as it shows the  best position of the detectors and also  how to optimize the system. Linearly sampled spectrum gives the best result as shown in figure~\ref{samp}.
}

\begin{figure}[htbp]
 \centering
  \includegraphics*[width=70mm]{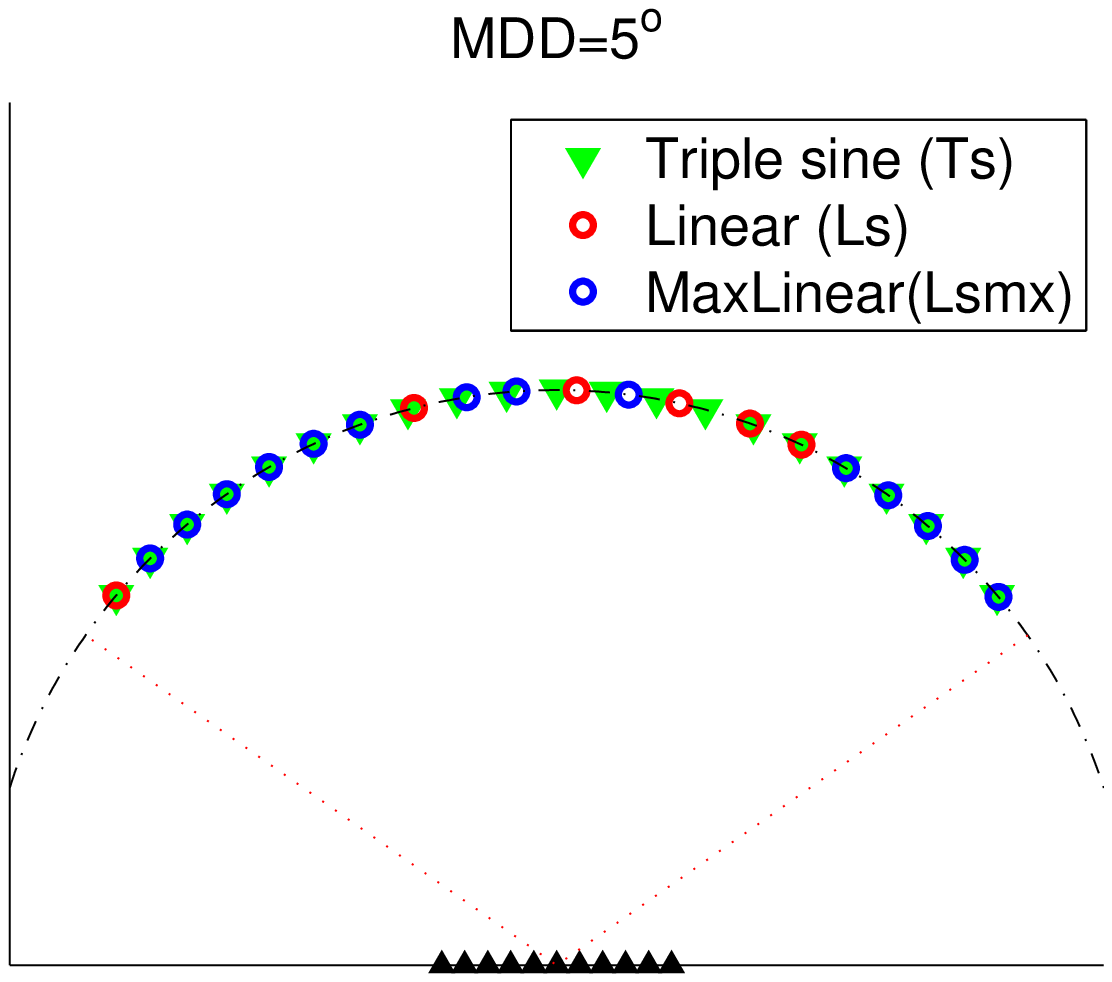} \\
    \includegraphics*[width=70mm]{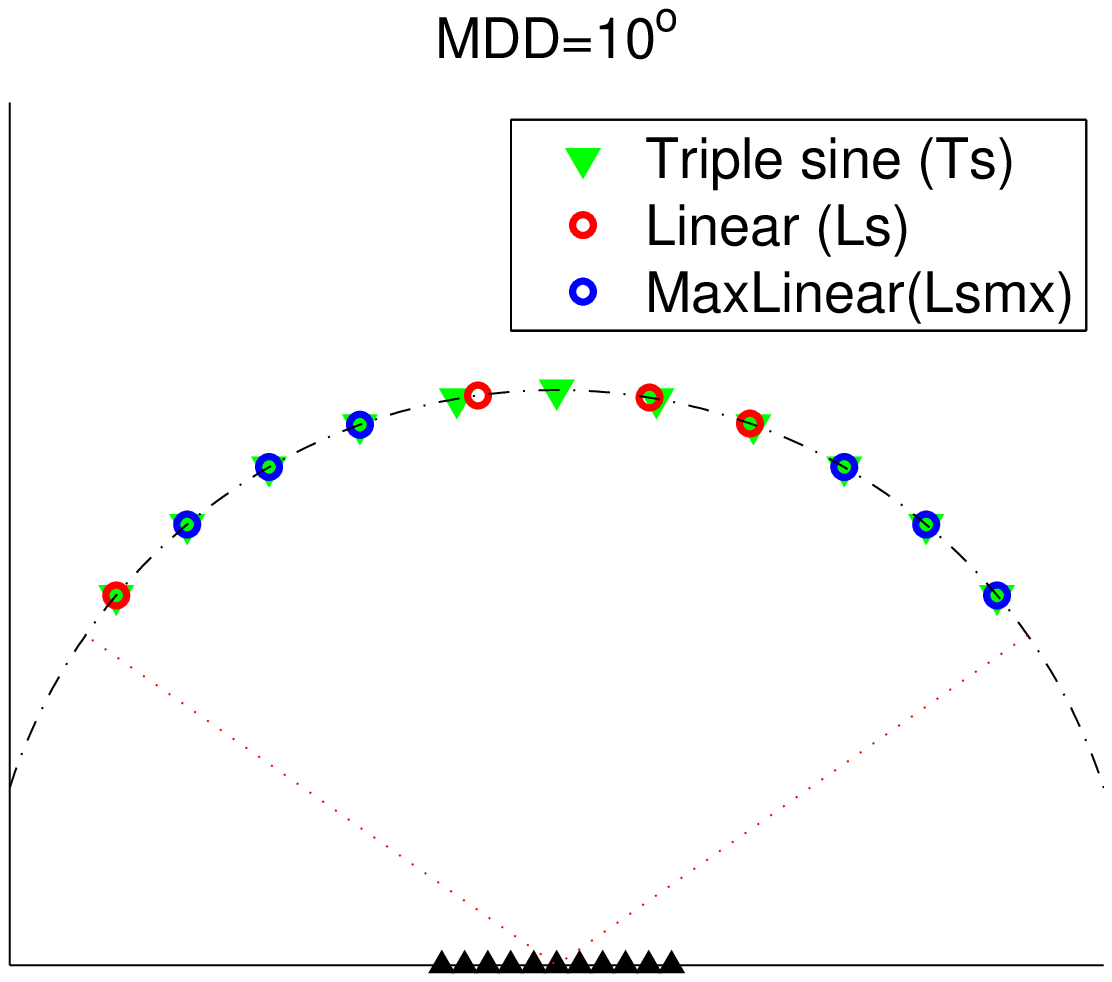}
  \caption{Detector position for linear sampling for a minimal detector distance of $5^o$ (top) and $10^o$ (bottom) .} 
   \label{lin12}
\end{figure}

{However, the linear sampling may not be practical to realize in a the real world. One needs to take into account the spatial size of the detectors (about 10 degrees in the case of E-203) and there is also a limit on the start and end points of detectors location (35-145 degrees for E-203). So linear sampling at a wide range of frequencies is impossible with this number of points. An investigation of how many linear sampling points can be used for a given angle difference between detectors shows that such physical constraints reduce strongly the number of detectors that can be used.
%For angle calculation and applying condition for first and final point was used  formula ~\ref{eq:lamb}.
Figure~\ref{lin12} shows examples of detector positions. The position of the red points is calculated using formula~\ref{eq:lamb} and blue are possible detector positions  which don't break the minimum detector distance (MDD) given on top of each plot.}

\begin{figure}[htbp]
 \centering
  \includegraphics*[width=90mm]{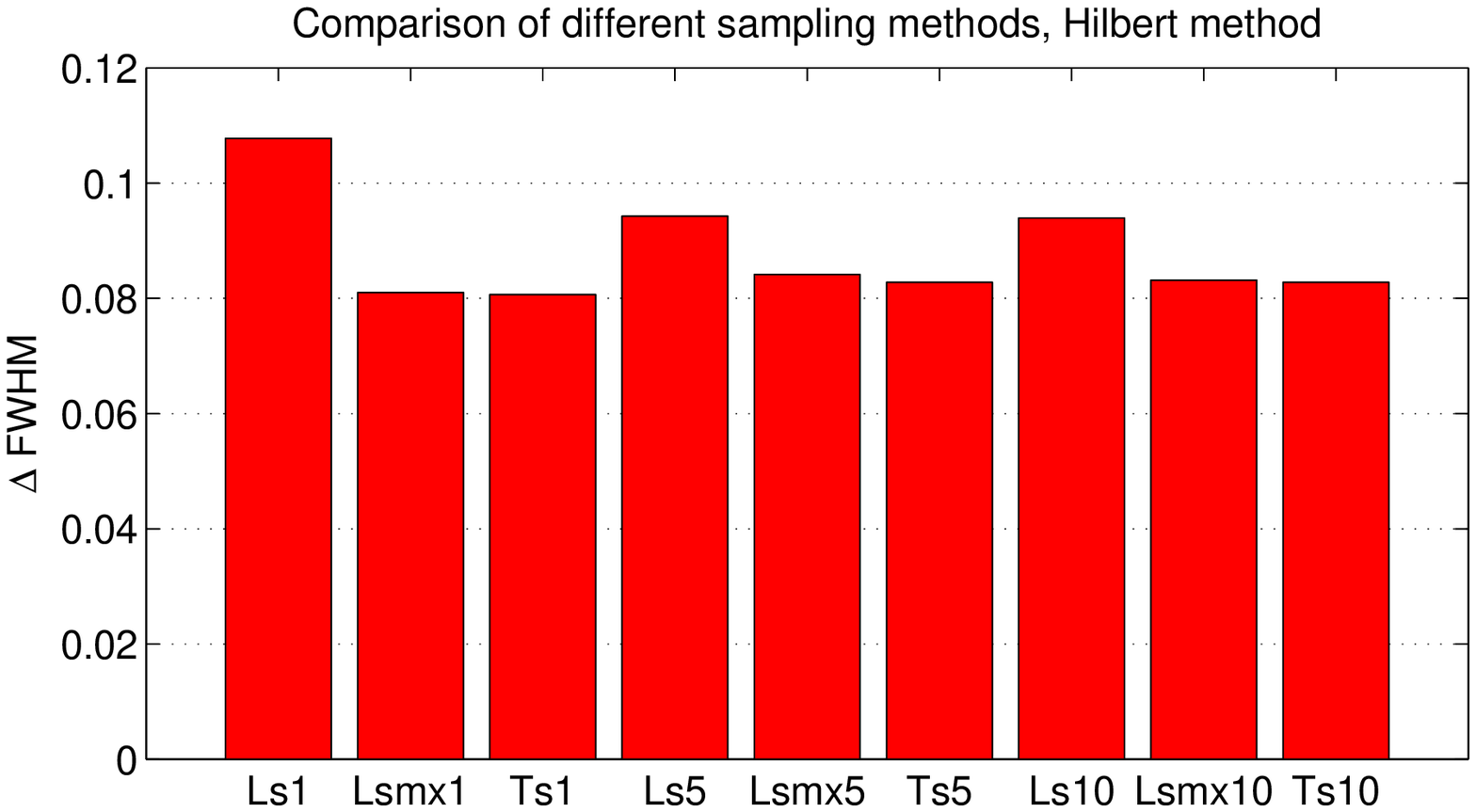} \\
  \includegraphics*[width=90mm]{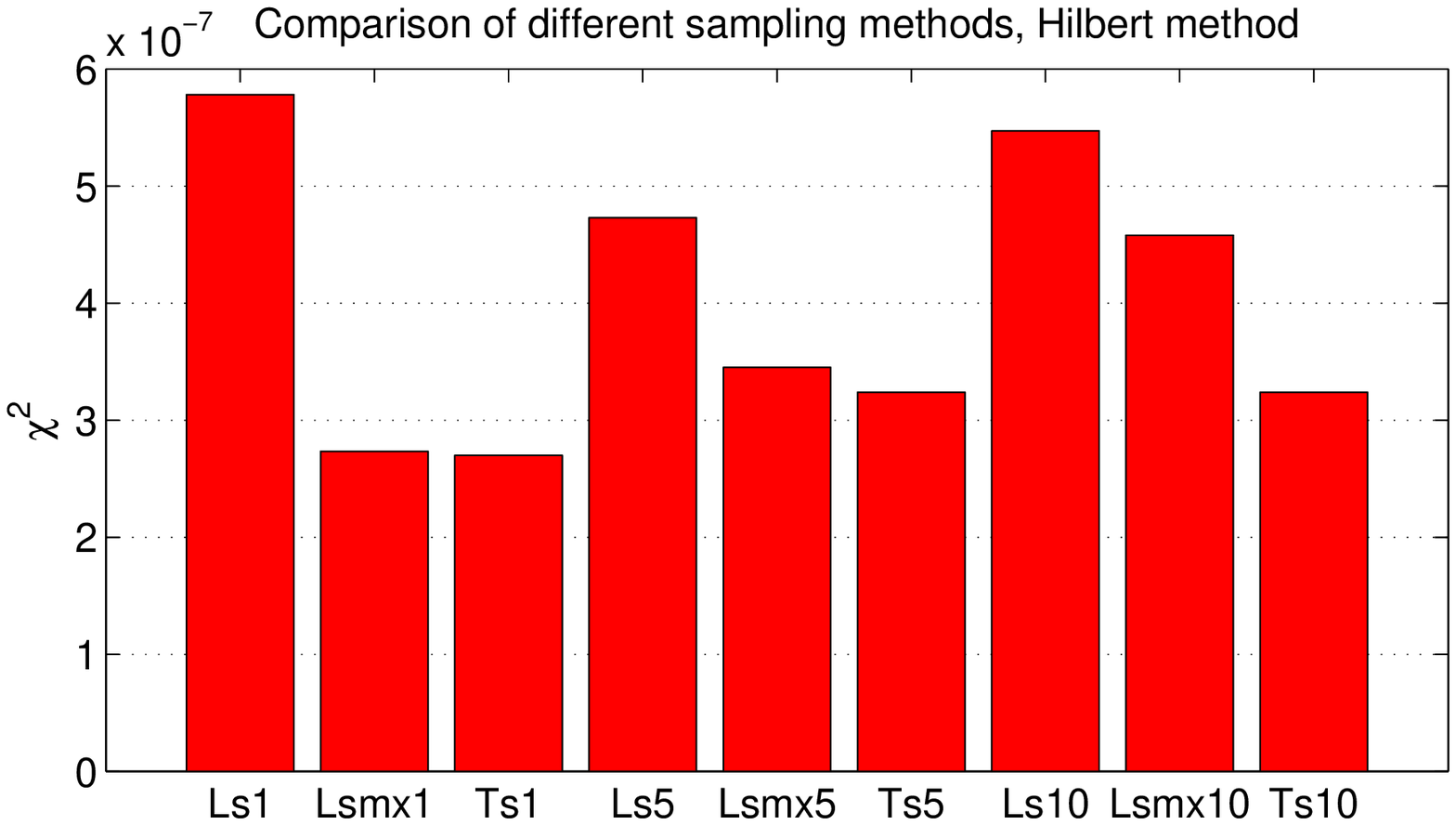} 
  \caption{Comparison of different sampling with number of MDD  with $\chi^2$ criterium (top) and $\Delta_{FWHM}$ (bottom). Ls is linear sampling with $1^o,5^o,10^0$ MDD and Ts is Triple sine sampling; mx mean the reconstruction use the maximum number of detectors (blue and red dots in figure \ref{lin12}).}
   \label{biglin}
\end{figure}

Figure~\ref{biglin} shows a comparison of the performances achieved with such positioning for different MDD. In each case the triple sine sampling (Ts) is better than the linear sampling (Ls) and close from the maximum number of detectors with linear sampling (Lsmx). 
%As the Lsmx configuration is physically impossible, 
So the Ts configuration is favored and will be used in the rest of this paper. The comparison between Ts1, Ts5 and Ts10 shows that reconstruction performances are limited by the MDD.

\begin{figure}[htbp]
 \centering
  \includegraphics*[width=70mm]{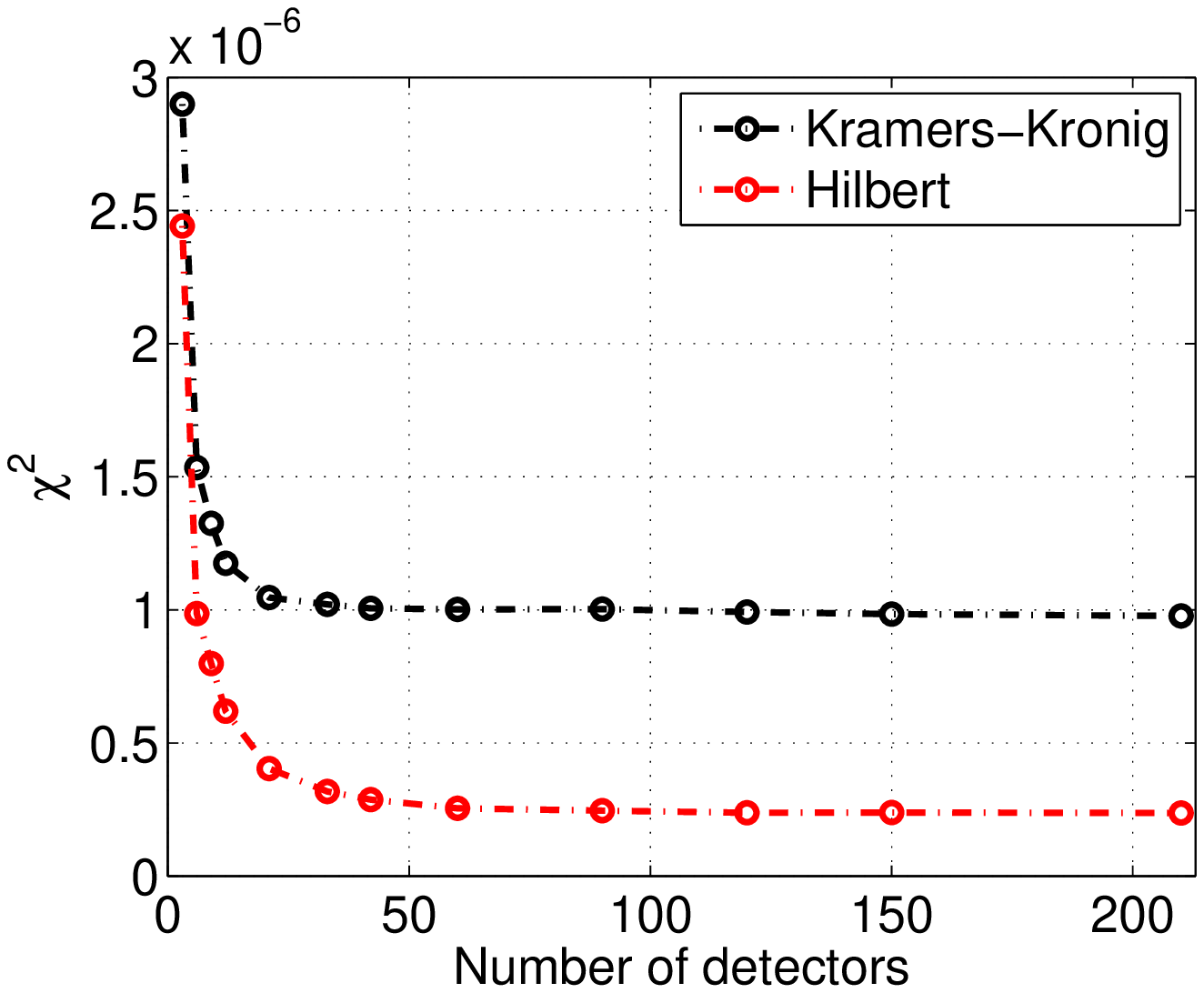}\\
   \includegraphics*[width=70mm]{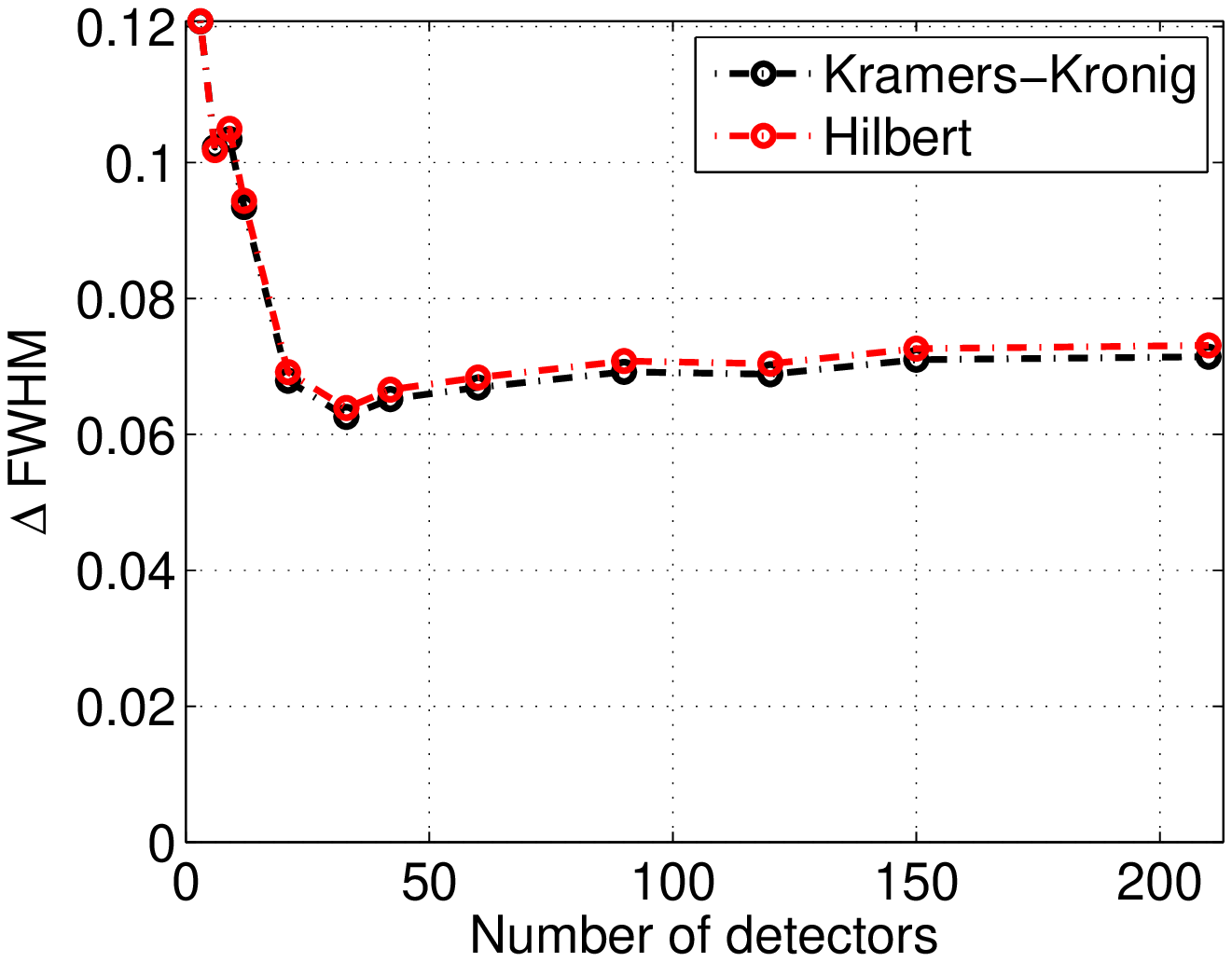}
  \caption{Effect of the sampling frequencies on the $\chi^2$ (top) and $\Delta_{FWHM}$ (bottom). }
   \label{sampling_chi2}
\end{figure}

%This section has been moved from elsewhere
The choice of 33 frequencies for the sampling of the spectrum was made to match the current layout used on E-203. However it is important to check if there is an optimum value. To perform this check we used the same simulations and the same simulated spectrum but sampled  with 3 to 140  points. The effect of changing the sampling frequencies on the  $\chi^2$ is shown in figure~\ref{sampling_chi2}. This study uses Triple sine sampling with 1000 profiles for each point and both reconstruction method.

It can be seen at figure \ref{sampling_chi2} that beyond about 33 sampling points the gain on the reconstructed $\chi^2$ is marginal.

%%%% Extrapolation / interpolations

After applying the sampling procedure the data need to be interpolated and extrapolated to have a larger number of points in the spectrum. Interpolation is done using Piecewise Cubic Hermite Interpolating Polynomial (PCHIP)~\cite{pchip}, as suggested in~\cite{VBthesis}.
The interpolation function must satisfy the following criteria: it must conserve the slope at the two endpoints (to have a continuous derivative) and respects monotonicity. PCHIP interpolation has been chosen as it matches these requirements. 

For low frequency extrapolation two methods have been investigated: Gaussian or Taylorian.

In the Gaussian method, we defined the extrapolation as follow:
\begin{equation}
\rho_{LF}(\omega)=Ae^{-(\omega-B)^2/2C^2}
\end{equation}
Where  $\rho_ {HF} (\omega)$ is the extrapolated spectrum at low frequency and the constants A, B, and C were chosen from the following conditions:
\begin{itemize}
\item $\rho_{LF}(0)=1$
\item $\rho_{LF}(\omega_0)=\rho(\omega_0)$
\item $\rho_{LF}'(\omega_0)=\rho'(\omega_0)$
\end{itemize}

The extrapolation relies in the fact that according to the central limit theorem in the time space the expected profile is Gaussian-like and in the frequency space it will also be Gaussian.

The other extrapolation method is based on Taylor expansion with the following definition:
\begin{equation}
\begin{split}
F(\omega)=\int_0^\infty dtS(t)e^{-i(\omega t)}
=\int_0^\infty dtS(t) \sum_{k=0}^{\infty}\frac{(-i\omega t)^k}{k!}=\\
= \sum_{k=0}^{\infty} \left(\frac{(-i\omega)^k}{k!} \int_0^\infty dtS(t)t^k\right)
=\sum_{k=0}^{\infty} \left(\frac{(-i\omega)^k}{k!} <t^k>\right)
\end{split}
\end{equation}

Approximation to the 4th order gives the following low frequency (LF) extrapolation:

\begin{equation}
\rho_{LF}=|F(\omega)|=\sqrt{A+B\omega^2+C\omega^4}
\end{equation}

\begin{figure}[htbp]
 \centering
  \includegraphics*[width=65mm]{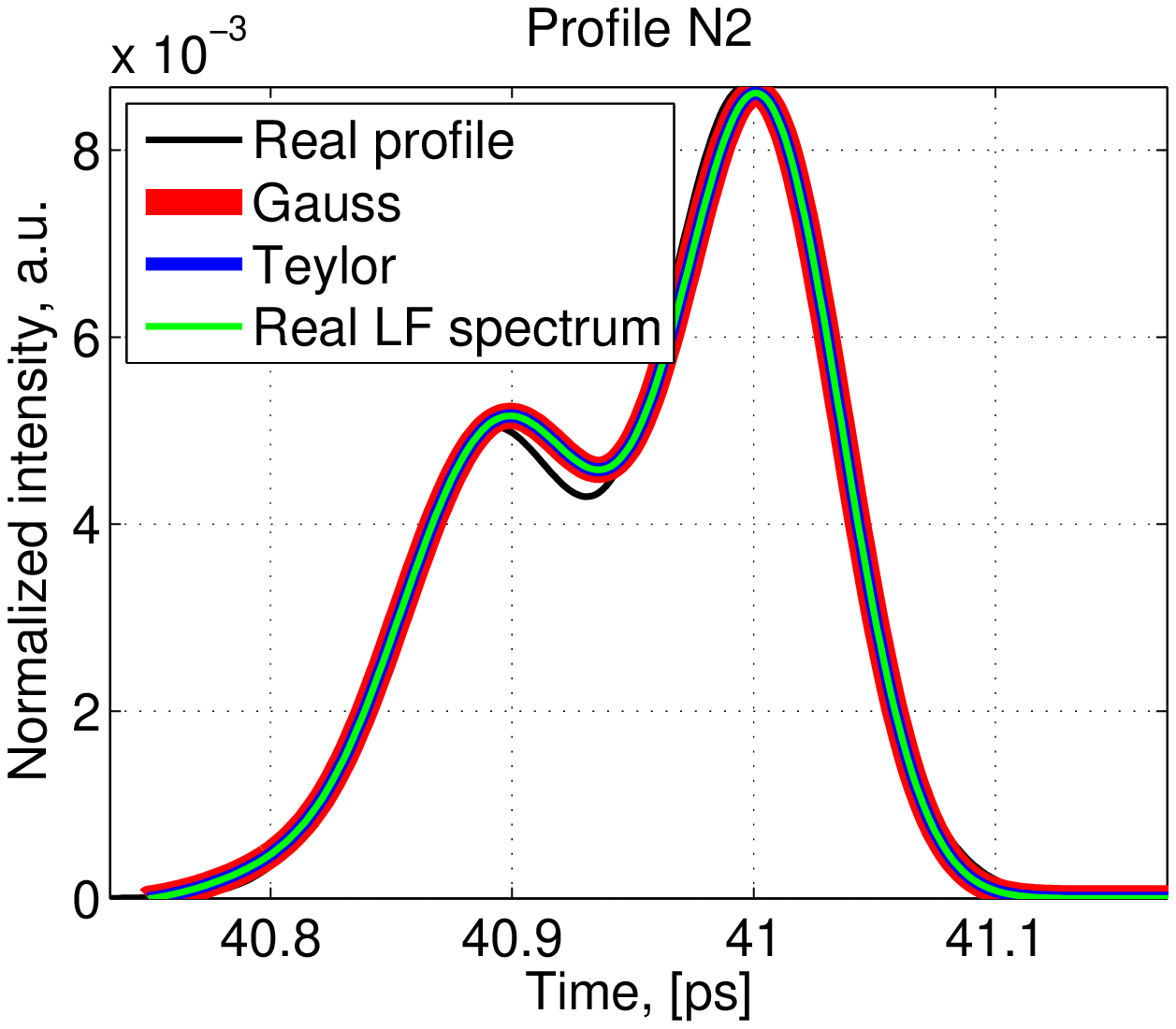}\\
  \includegraphics*[width=65mm]{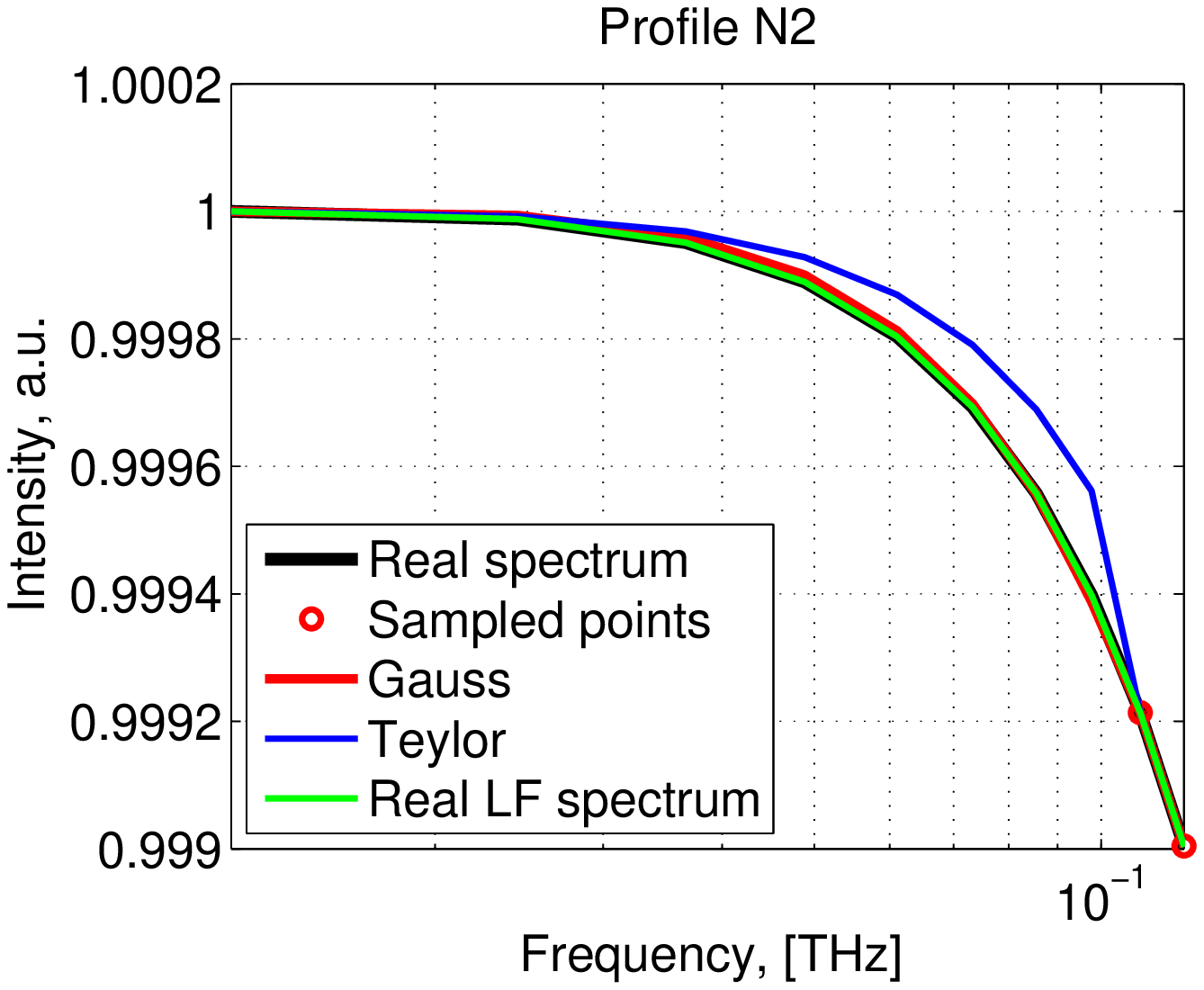}\\

    \caption{Comparison of different LF extrapolation: example of spectrum (top) and  profile (bottom) and histogram with mean $\chi^2$ for each method (bottom). Gaussian and Taylorian methods are described in the text. "Real LF spectrum" means that the real LF spectrum is used. For this simulation the Hilbert method of phase recovery and $A\omega^B$ high frequency extrapolation were used.}
   \label{lf}
\end{figure}

\begin{figure}[htbp]
 \centering

  \includegraphics*[width=65mm]{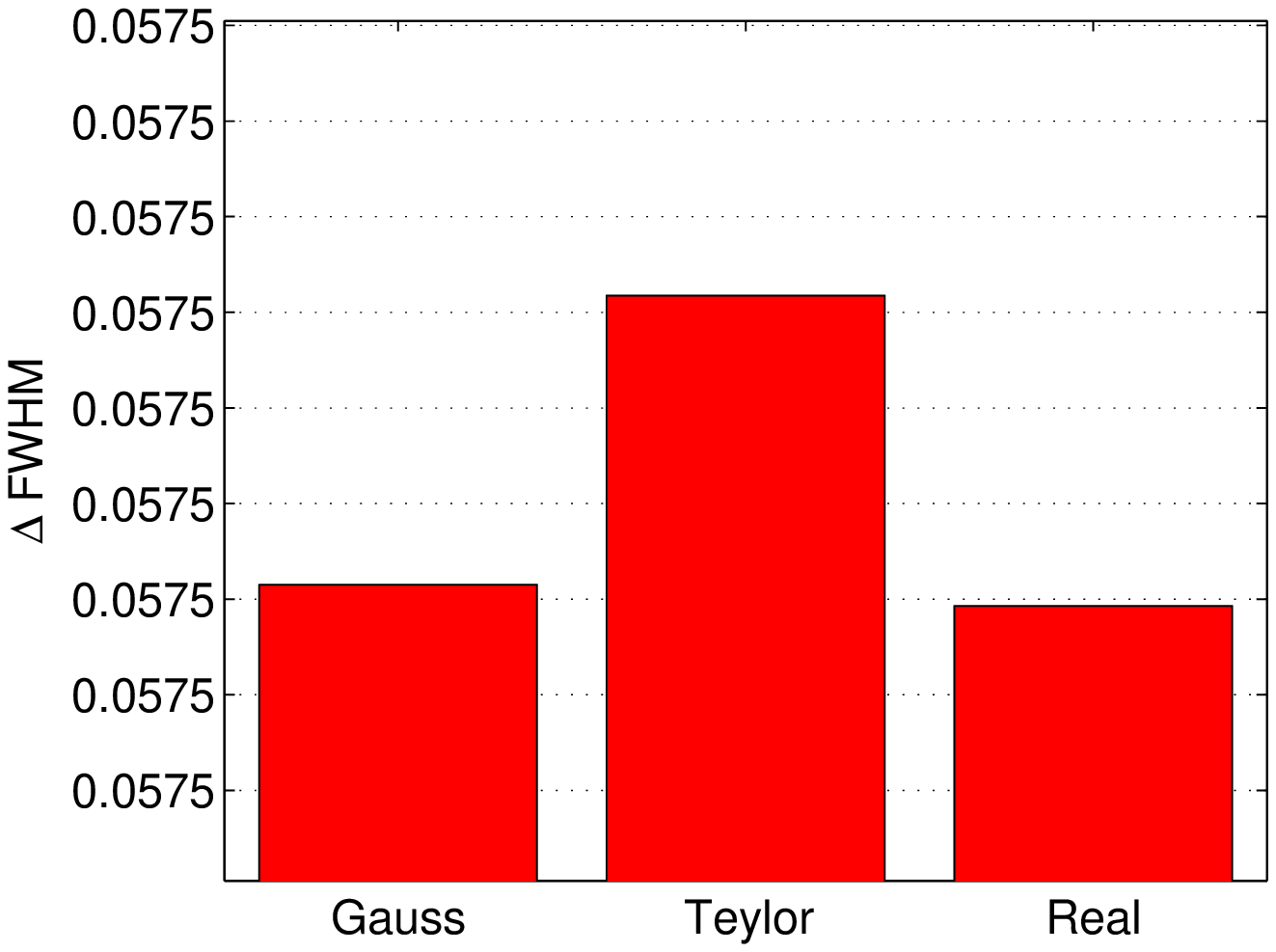}
    \includegraphics*[width=65mm]{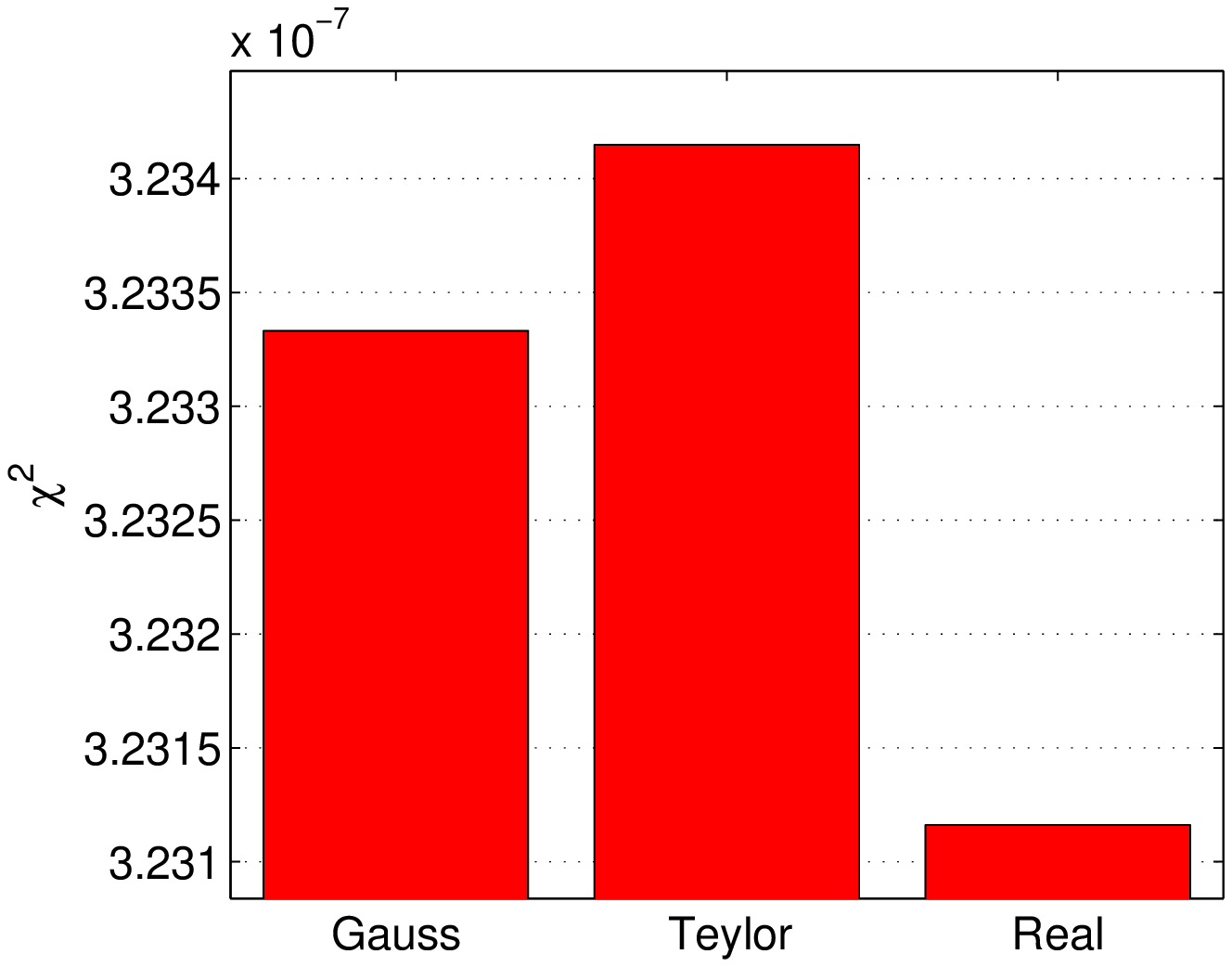}
    \caption{Comparison of different LF extrapolation: histogram with mean $\chi^2$ for each method (top) and $\Delta_{FWHM}$ (bottom).  }
   \label{lf2}
\end{figure}

Conditions for the constants A, B and C  are the same. Comparison of different LF extrapolation can be found in figure~\ref{lf} and the performances of these methods in figure~\ref{lf2}.

In the rest of this paper we used the Gaussian method.

Several high frequency (HF) extrapolation methods were also tested. The most common~\cite{VBthesis,DESYthesis} is :
\begin{equation}
\rho _ {HF} (\omega)=A\omega^{-4},
\end{equation}
 where $\rho_ {HF} (\omega)$ is the extrapolated spectrum at high frequency and $A=\rho_f \omega_f^{4} $, where $\rho_f$ is spectrum value of final point $\omega_f$.

The second method uses the same consideration as in Lai and Sievers~\cite{LaiS}: 
Assuming that the bunch size is finite with two end points at $z=0$ and at $z=\sigma_z$ then the longitudinal charge distribution ($S$) must follow $S(0)=S(\sigma_z)=0$. 
An integration by parts gives~:
\begin{multline}
F(\omega)=\int_0^\infty dzS(z)e^{i(\frac{\omega}{c})z}=\\
=\frac{S(z)}{i\frac{\omega}{c}} 
e^{i(\frac{\omega}{c})z} \Big|_0^{\sigma_z}-\frac{S^{\prime}(z)}{\Big(i\frac{\omega}{c}\Big)^2} 
e^{i(\frac{\omega}{c})z} \Big|_0^{\sigma_z}+\ldots
\end{multline} 
The first term vanishes because of the boundary conditions, so for large $\omega$, $F(\omega)$ is proportional to $\omega^{-2}$ and two conditions have to be matched~:
\begin{itemize}
\item $\rho_{HF}(\omega_{fmax})=\rho(\omega_{fmax})$
\item $\rho_{HF}'(\omega_{fmax})=\rho'(\omega_{fmax})$
\end{itemize}
where $\omega_{fmax}$  is the last sampled point  of the spectrum. To satisfy the boundary condition two constants are needed, giving a two-terms extrapolation~:
\begin{equation}
\rho_{HF}(\omega)=A\omega^{-2}+B\omega^{-3}
\end{equation}
or  extrapolation with degree of frequency as free parameter:
\begin{equation}
\rho_{HF}(\omega)=A\omega^B
\end{equation}

where the A and B -- coefficients which are calculated from the last data samples and the boundary conditions as follow:
\begin{itemize}
\item $B={\rho_{HF}'(\omega_{fmax}) \omega_{fmax}}/{\rho(\omega_{fmax})}$
\item $A={\rho(\omega_{fmax})}/{\omega_{fmax}^B}$
\end{itemize}

  The requirement of finite bunch size requires $B\le2$, so in the case where the fit gives $B>-2$ we use $B=-2$.
Two other extrapolation methods also have been investigated:
\begin{itemize}
\item $\rho_{HF}(\omega_f)= 0 $ for $ \omega_f >  \omega_{fmax}$
\item $\rho_{HF}(\omega_f)= \rho_{real}(\omega_f) $ for $ \omega_f >  \omega_{fmax} $ where $\rho_{real}$ is the real spectrum.
\end{itemize}

\begin{figure}[htbp]
 \centering
  \includegraphics*[width=65mm]{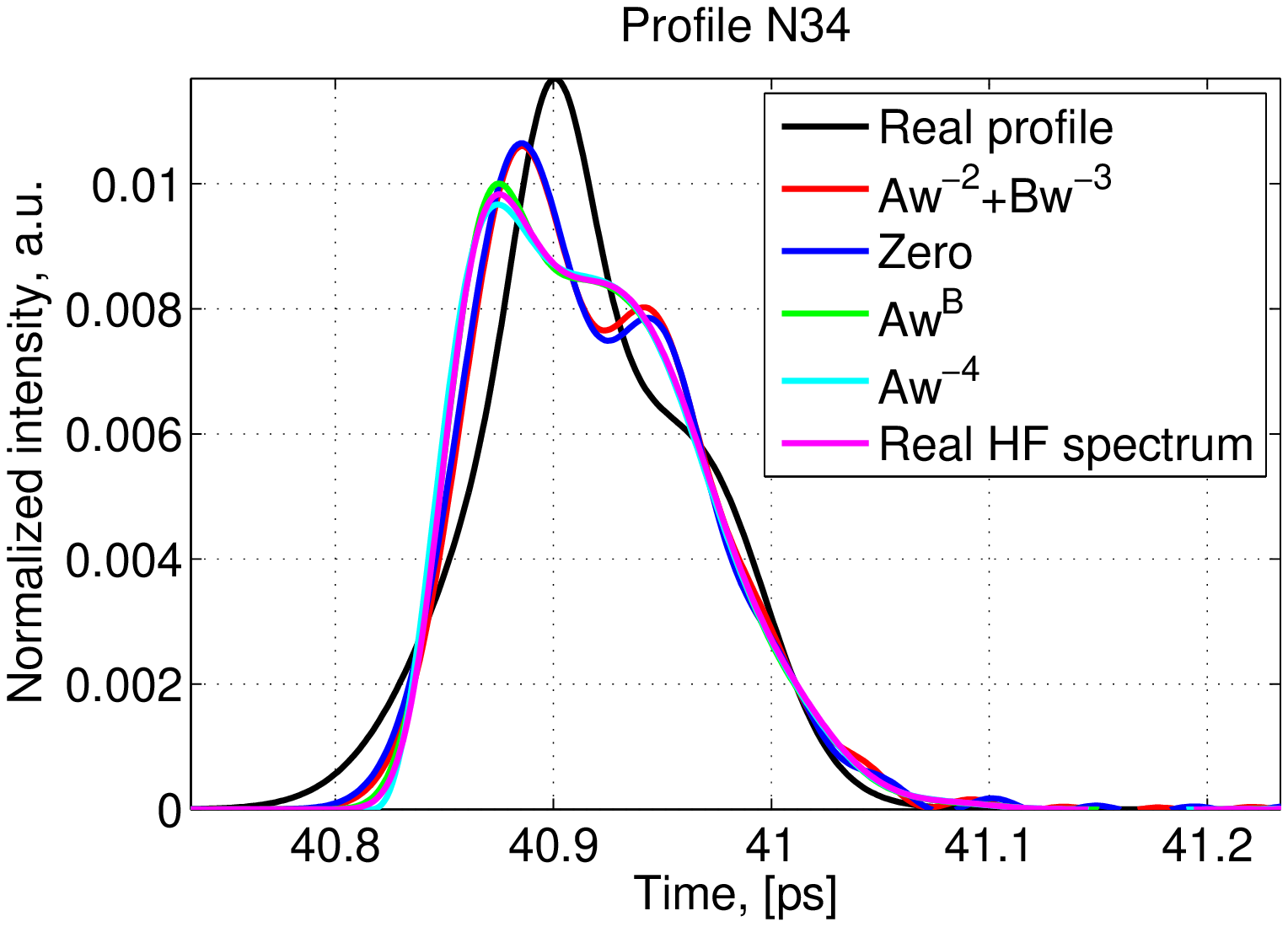}\\
  \includegraphics*[width=65mm]{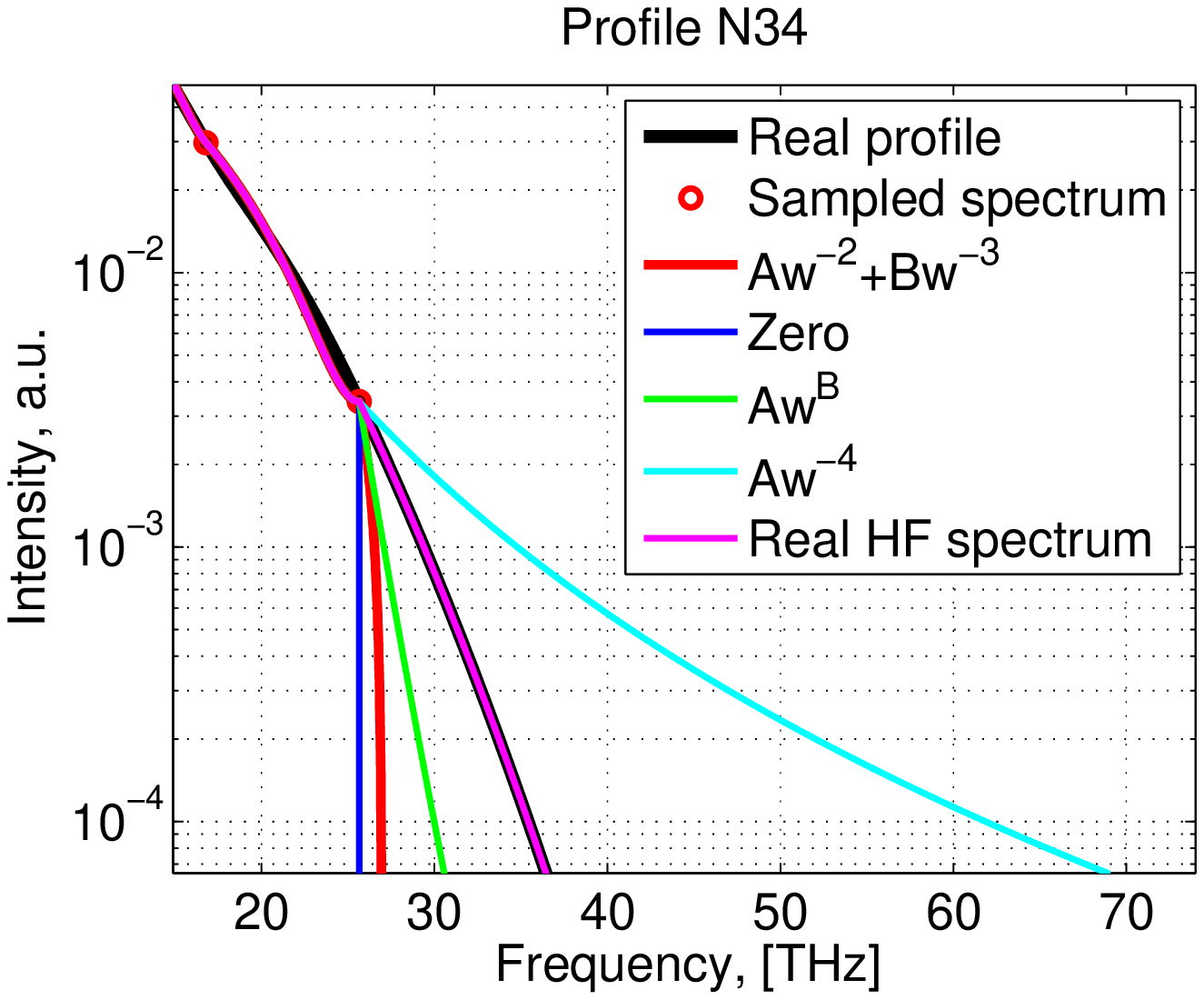}\\

    \caption{Comparison of different HF extrapolations~: example of spectrum  (top) and profile (bottom). For these simulations the Hilbert reconstruction method of phase recovery and Gaussian LF extrapolations were used.}
   \label{hf}
\end{figure}

\begin{figure}[htbp]
 \centering
    \includegraphics*[width=65mm]{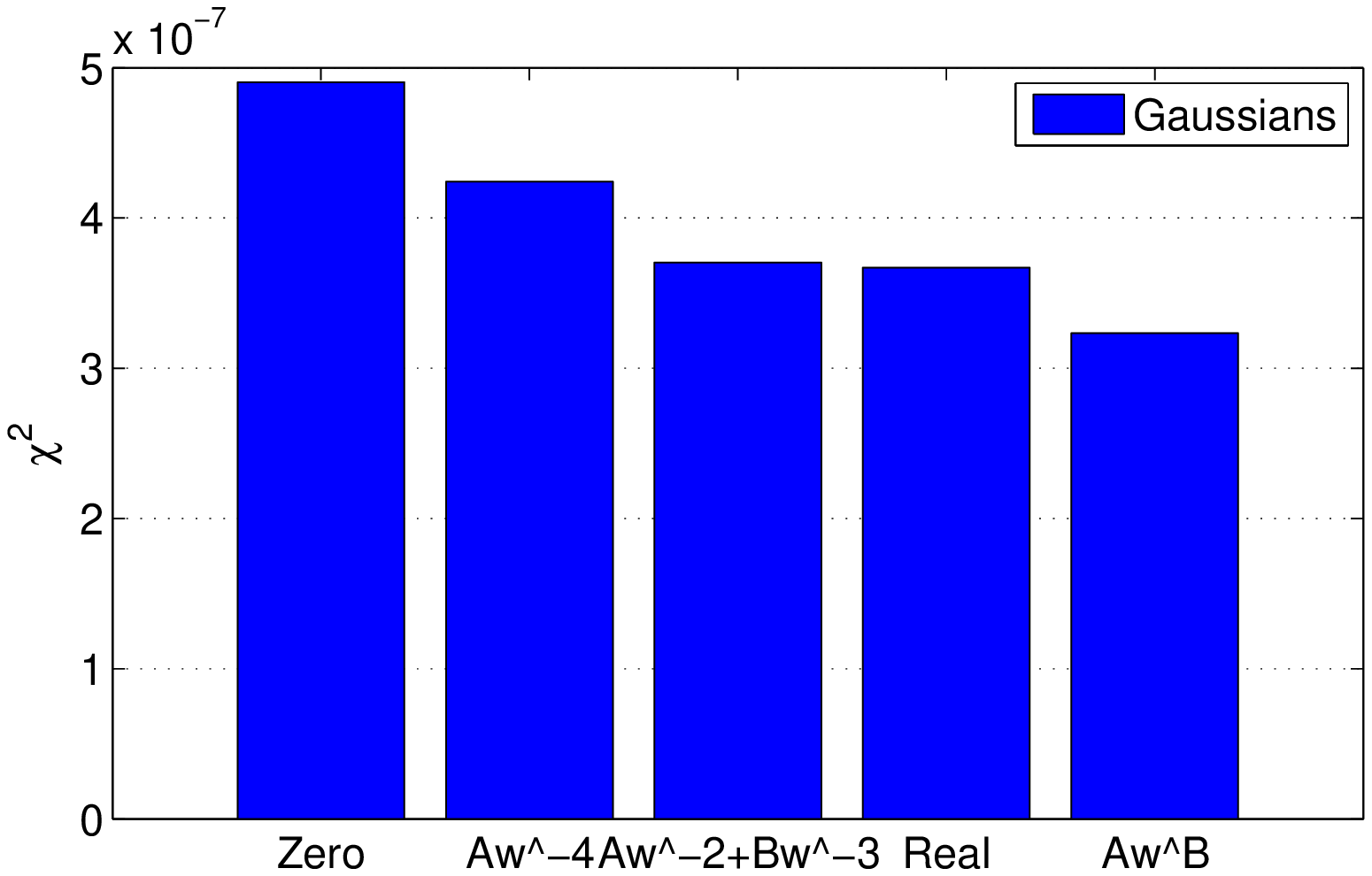}\\
        \includegraphics*[width=65mm]{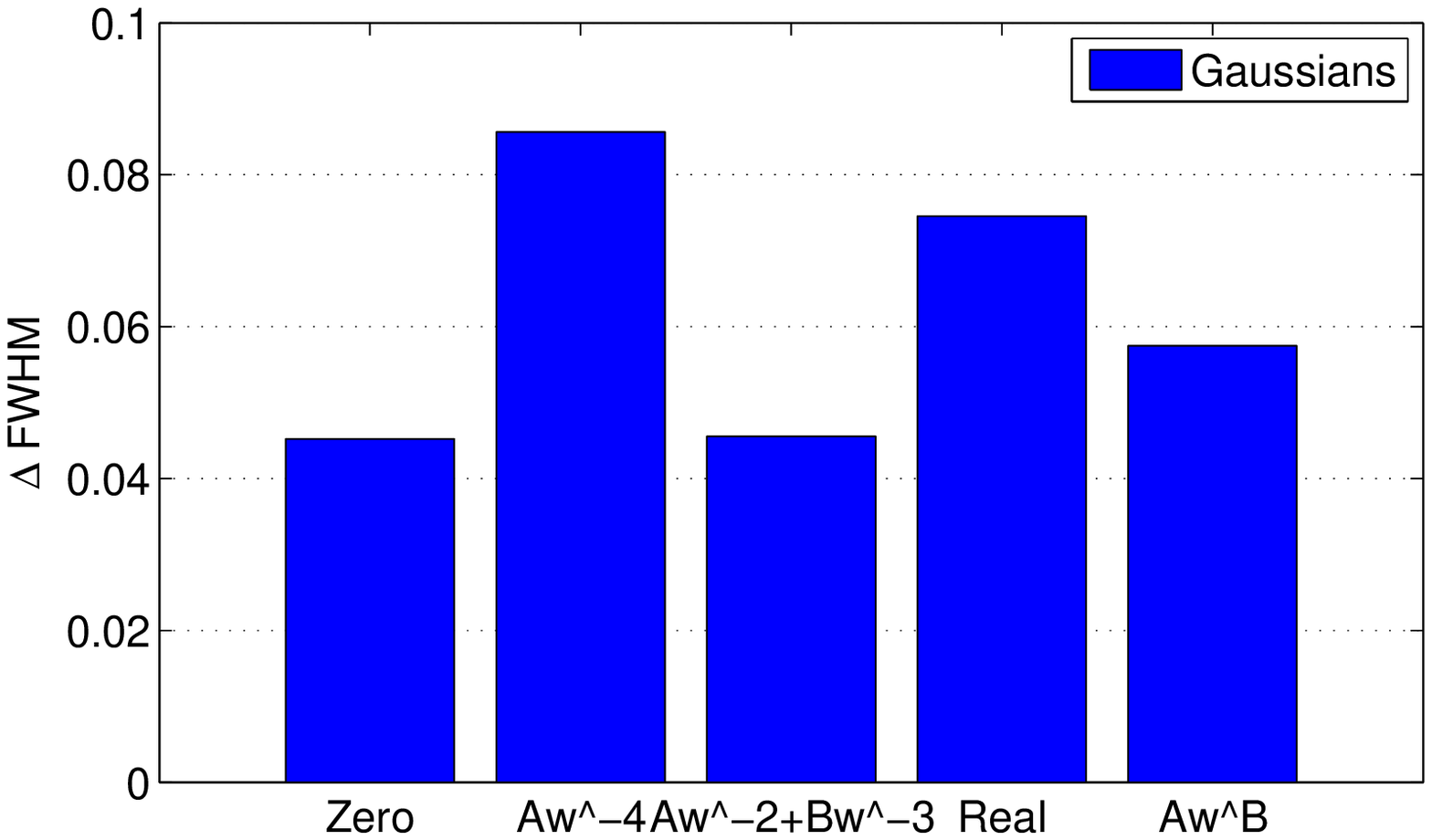}\\

    \caption{Comparison of different HF extrapolation for Gaussian~: histogram with mean $\chi^2$ (top) and $\Delta_{FWHM}$ (bottom).}
   \label{hf2}
\end{figure}

These HF extrapolation methods are compared in figure~\ref{hf} and~\ref{hf2}.\par
Thus, by virtue of the above arguments and simulations, it's naturally to choose the high-frequency extrapolation by power function.


%\begin{figure}[htbp]
% \centering
%        \includegraphics*[width=65mm]{new203/HFLorenz.eps}\\
%          \includegraphics*[width=65mm]{rev4/hflor.eps}
%        \caption{Comparison of different HF extrapolation for Lorenzians~:histogram with mean $\chi^2$ (top) and $\Delta_{FWHM}$ (bottom).}
%   \label{hf3}
%\end{figure}

%\textbf{XXX If you show Lorenzian profiles for HF extrapolation why don't you also show it for LF extrapolation? XXX}

\section{Study of the reconstruction performance}

\begin{figure}[htbp]
 \centering
  \includegraphics*[width=65mm]{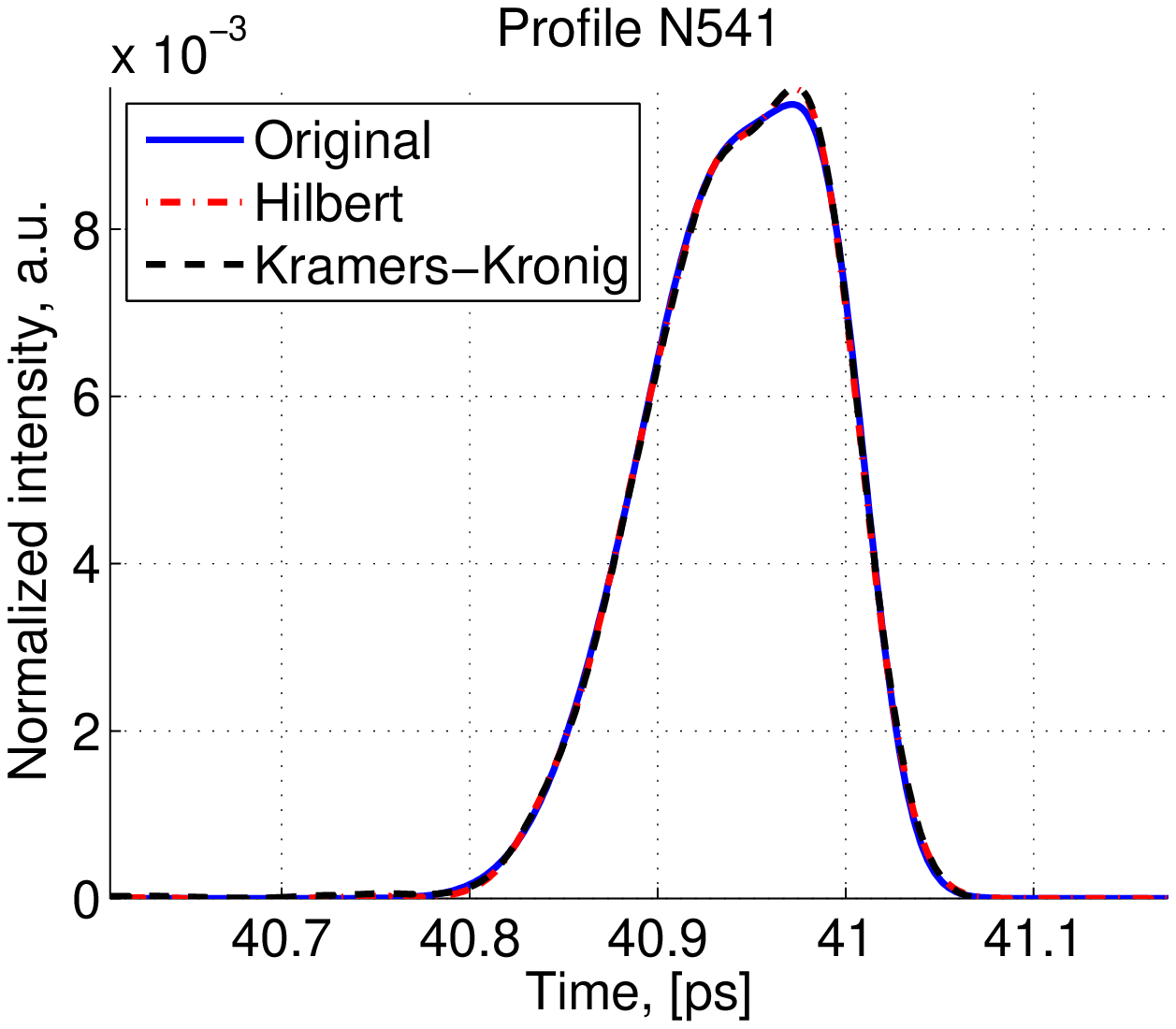}\\
  \includegraphics*[width=65mm]{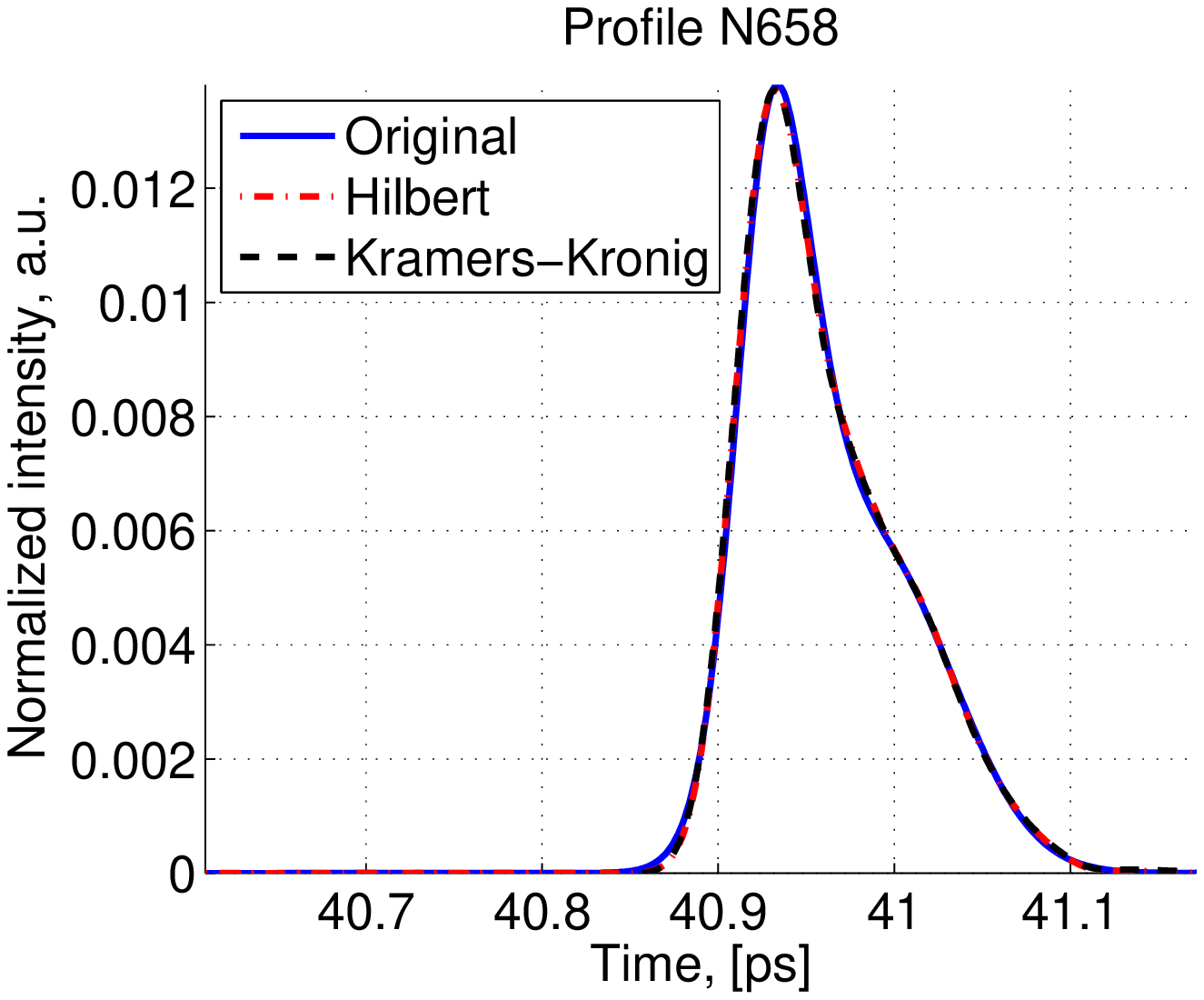}\\
   \includegraphics*[width=65mm]{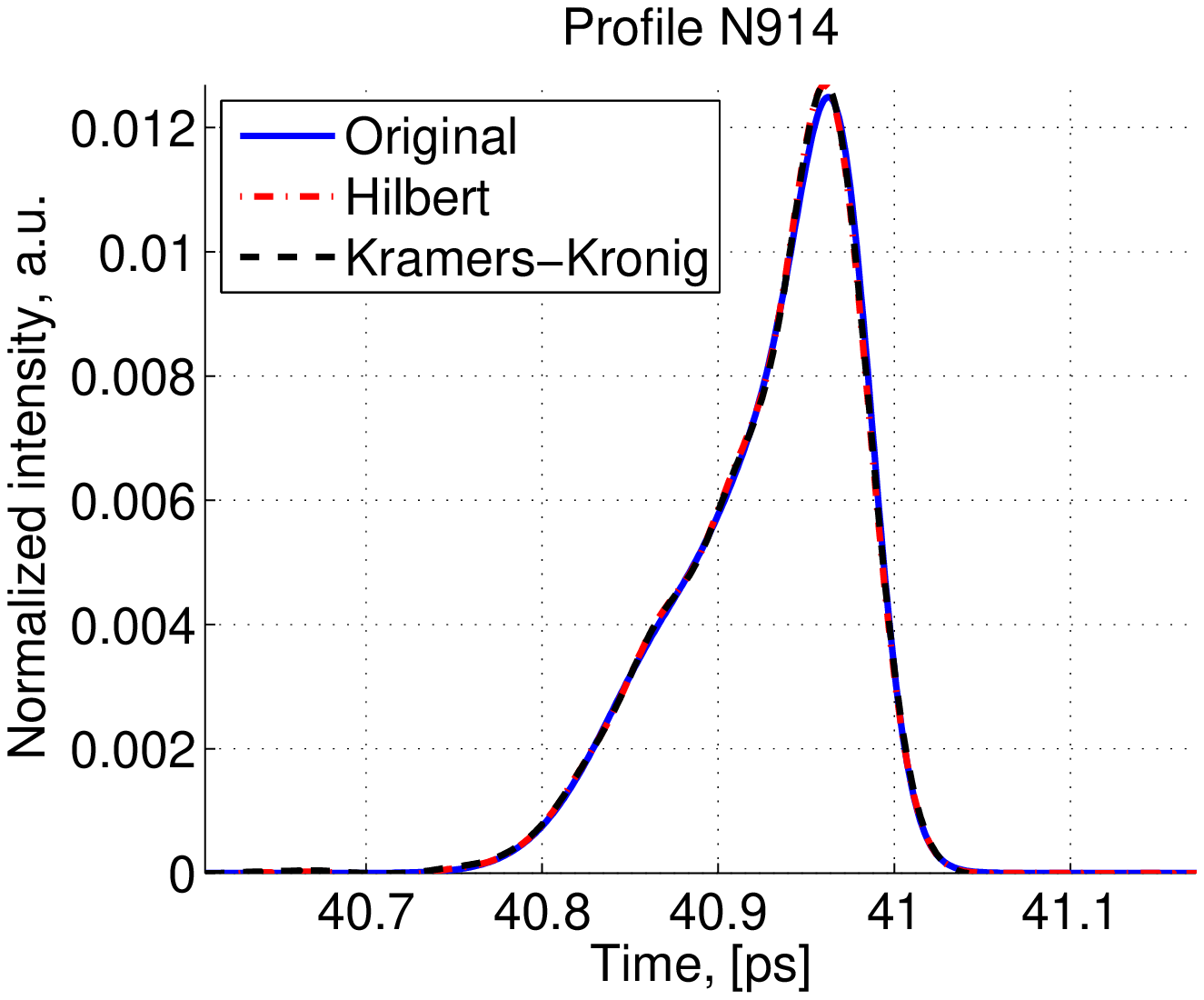}
  \caption{Examples of well reconstructed profile. The original profile is in blue and the profiles reconstructed with the Hilbert transform and the full Kramers-Kronig procedures are in red and black respectively.}
   \label{good_profiles}
\end{figure}

\begin{figure}[htbp]
 \centering
  \includegraphics*[width=65mm]{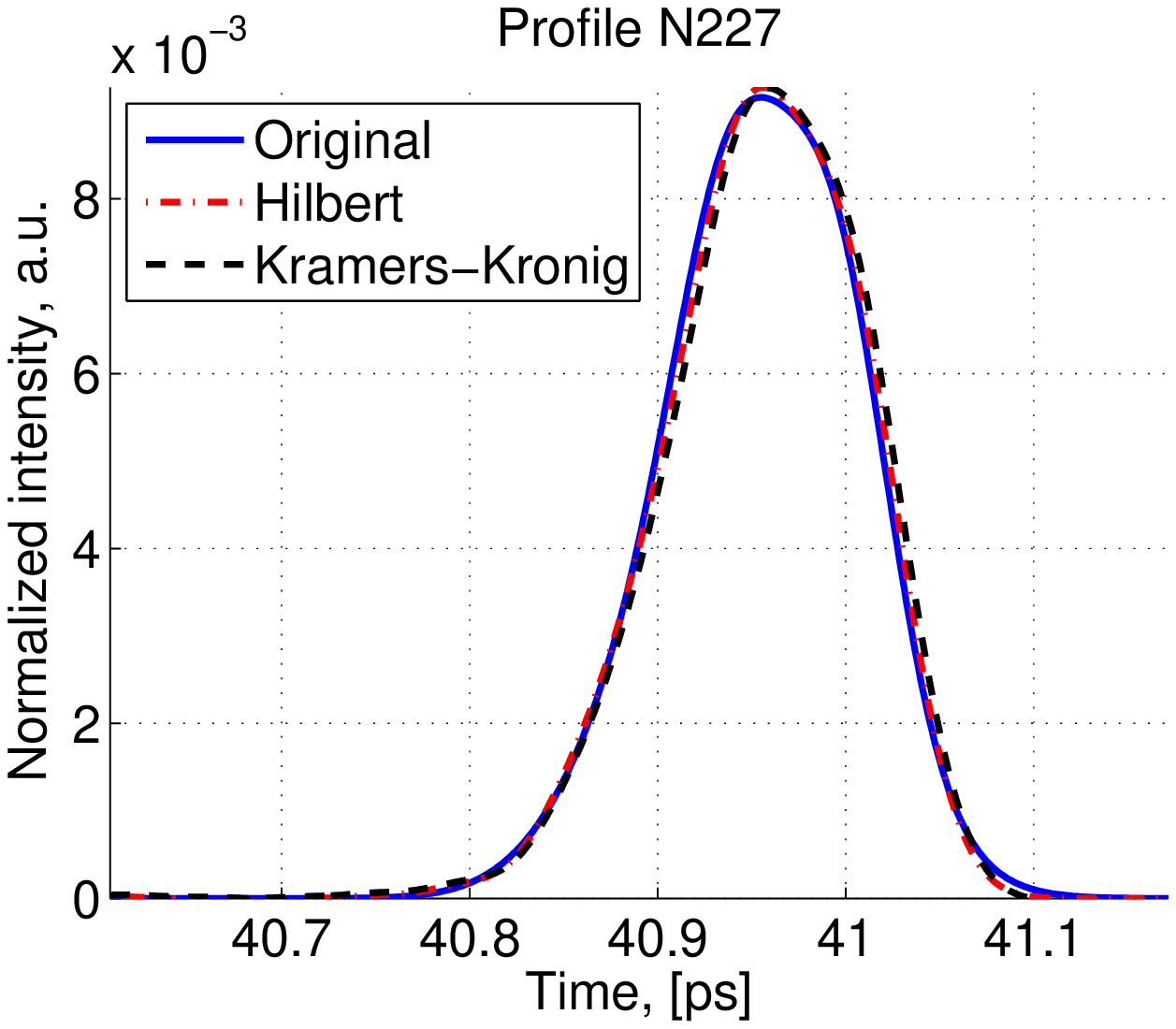}\\
   \includegraphics*[width=65mm]{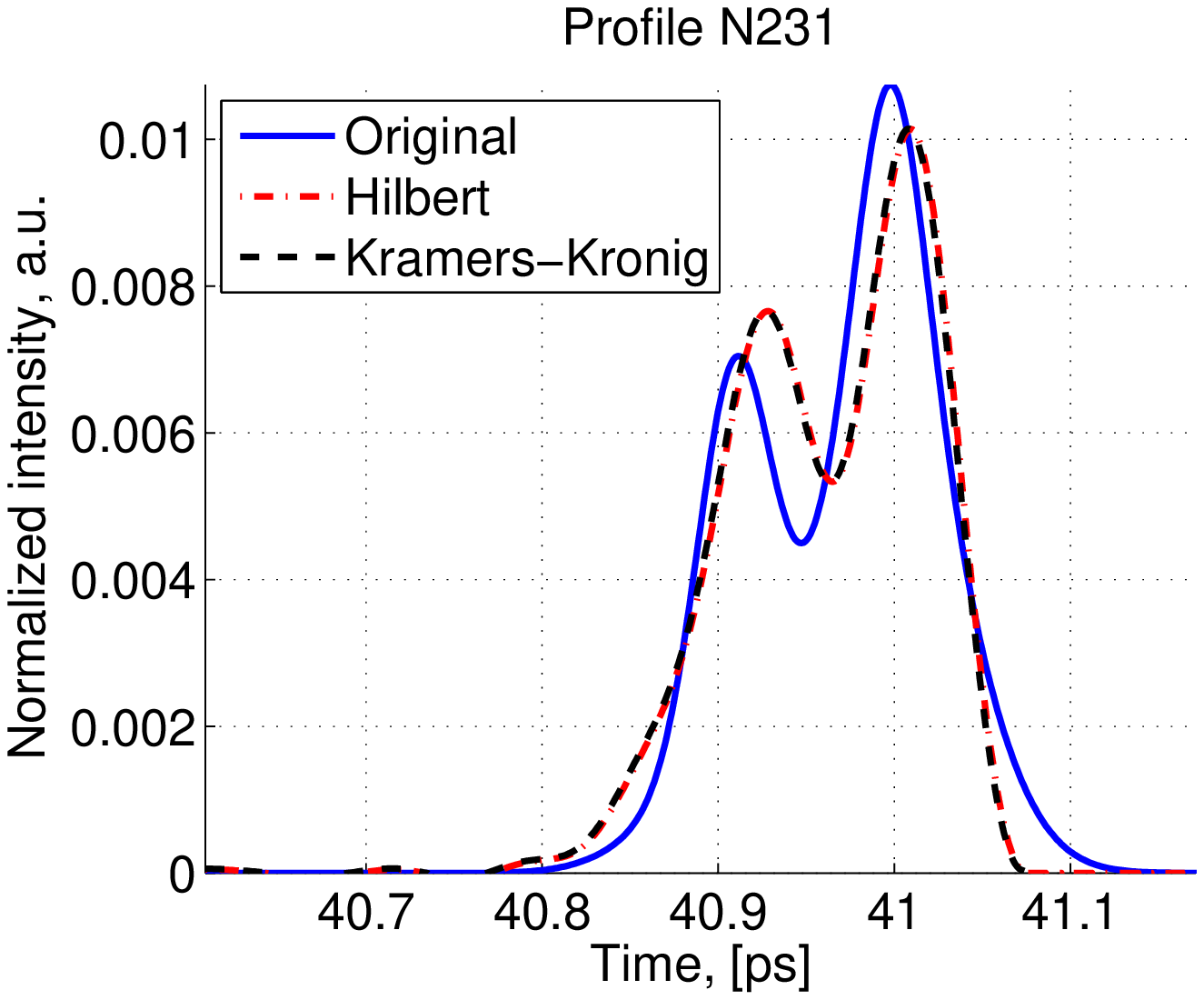}\\
    \includegraphics*[width=65mm]{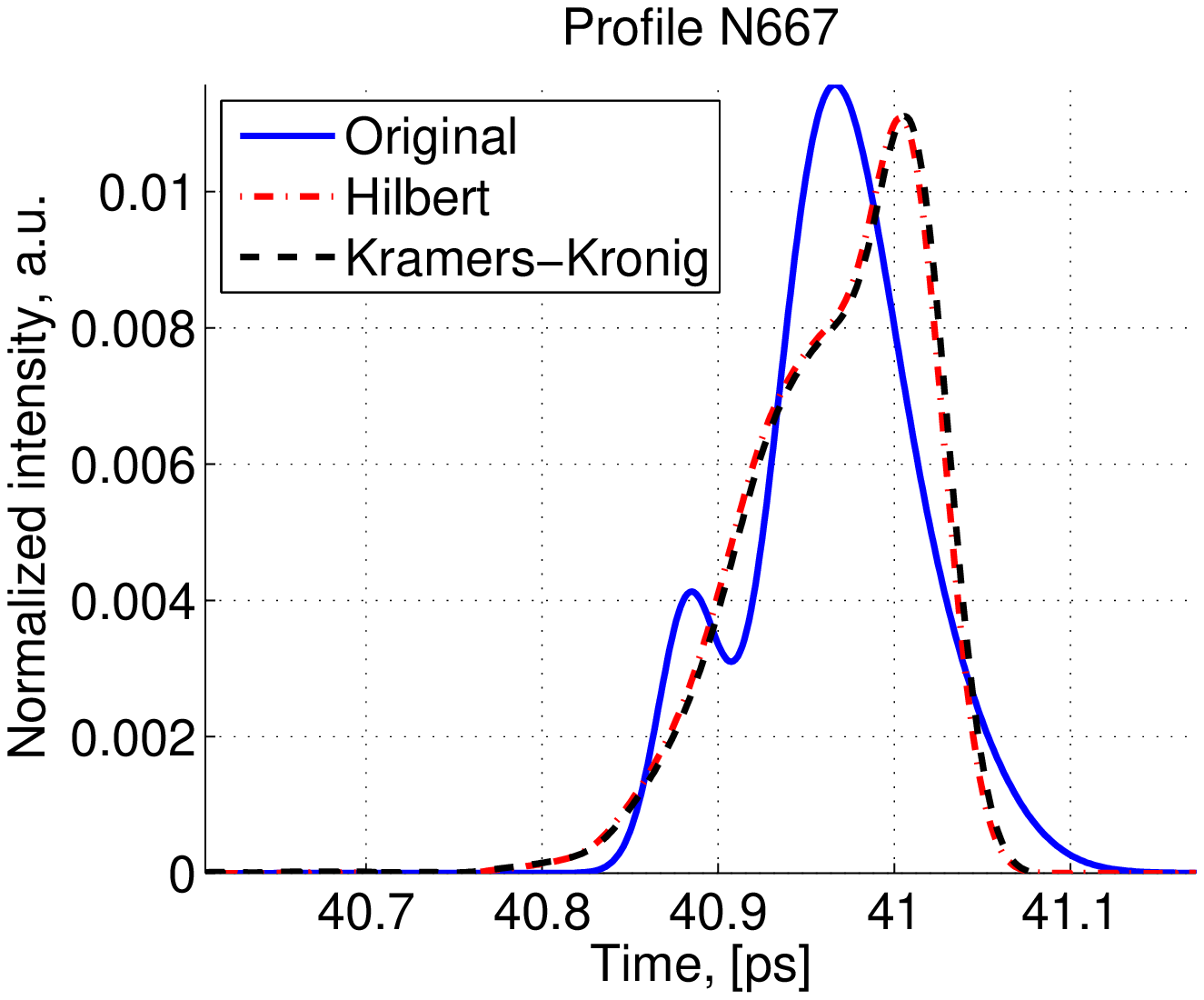}\\
  \caption{Example of poorly reconstructed profile.  The original profile is in blue and the profiles reconstructed with the Hilbert transform and the full Kramers-Kronig procedures are in red and black respectively.}
   \label{bad_profiles}
\end{figure}

After applying extrapolation and interpolation, the spectrum recovery is completed. Then we used  different reconstruction techniques to reconstruct the original profile. For each reconstruction method some profiles are very well reconstructed whereas some other are not so well reconstructed. Examples of well reconstructed profiles are shown in figure~\ref{good_profiles} and examples of poorly reconstructed profile are shown in figure~\ref{bad_profiles}.

\begin{figure}[htbp]
 \centering
  \includegraphics*[width=70mm]{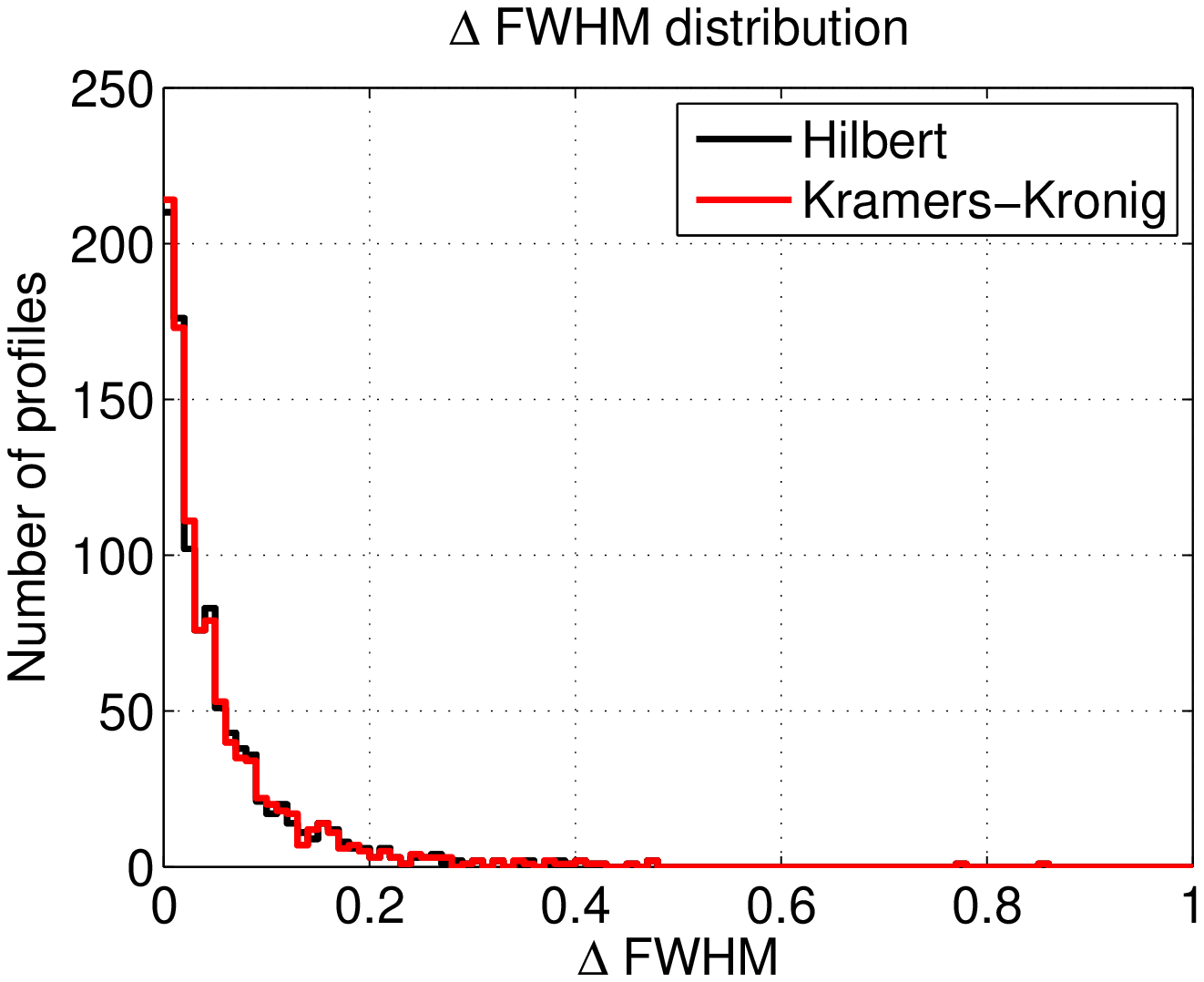} \\
  \includegraphics*[width=70mm]{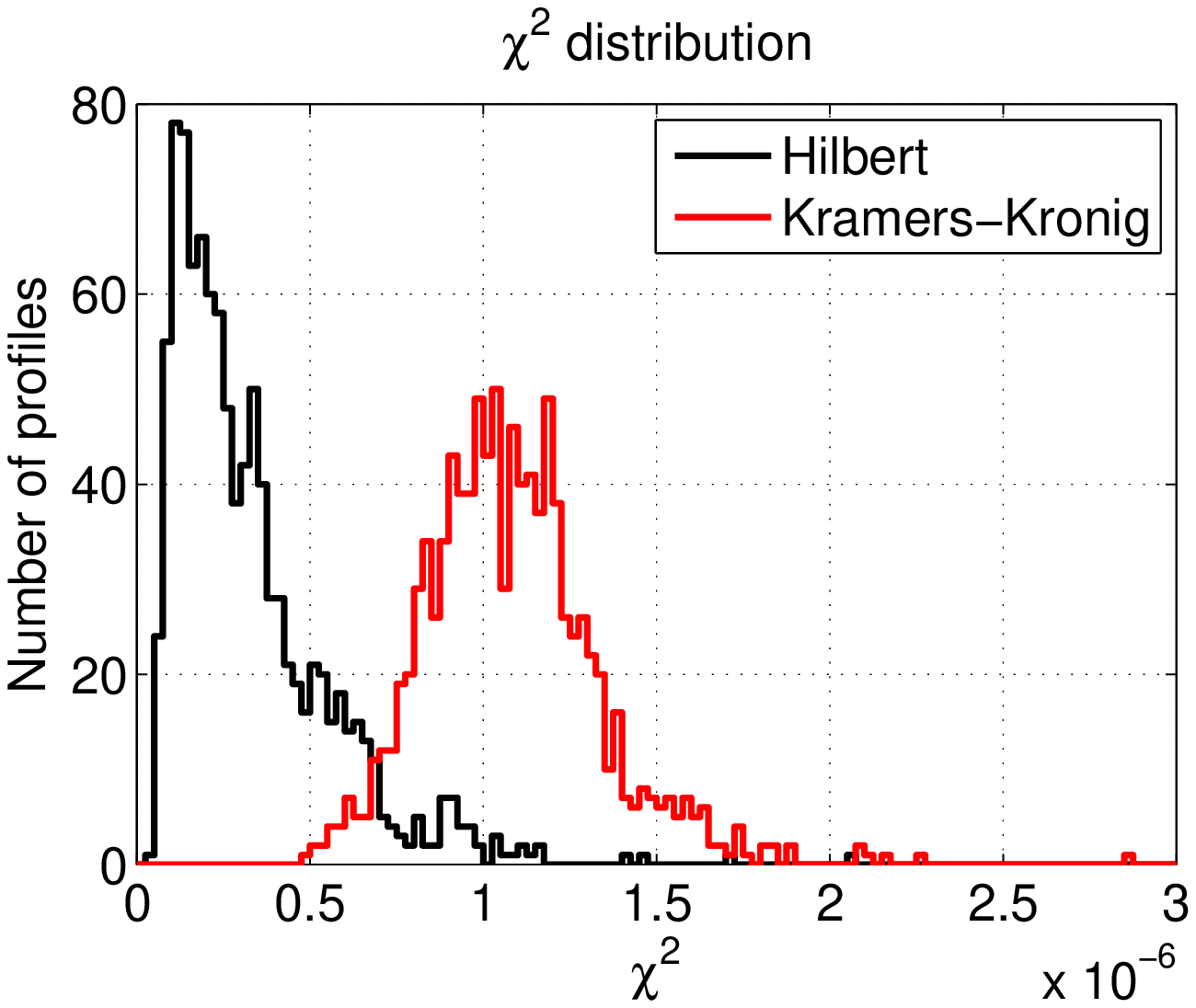} 
  \caption{{$\Delta_{FWHM}$  (top)  and $\chi^2$ (bottom) distribution of 1000 simulations reconstructed using the Hilbert transform  method (black line) and Kramers-Kronig reconstruction method (red line).  }}% VH change name of picture and unite with other
   \label{profiles_stats_hilbert}
\end{figure}

The  $\Delta_{FWXM}$ and  $\chi^2$ distribution of the 1000 simulations which were made and then reconstructed using the Hilbert transform method and Kramers-Kornig reconstruction are shown in figure~\ref{profiles_stats_hilbert}. There is a good concordance in FWHM between  two methods indicating that they are both good at finding the bunch length. However, the Hilbert method gives lower $\chi^2$ indicating that this method is better at reconstruction of the bunch profile.

\begin{figure}[htbp]
 \centering
  \includegraphics*[width=100mm]{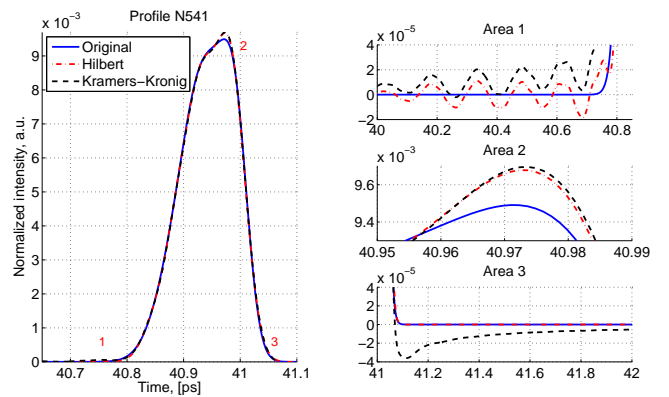}  
  \caption{Example of reconstructed profile with zooms on the peak and tails. One can see that the profile reconstructed using the Kramers-Kronig method has a negative component. This will dominate the final $\chi^2$ and explains why the $\chi^2$ obtained by this method is higher as shown in figure~\ref{profiles_stats_hilbert}.}% VH change name of picture and unite with other
   \label{expKK}
\end{figure}

\begin{figure}[htbp]
 \centering
  \includegraphics*[width=70mm]{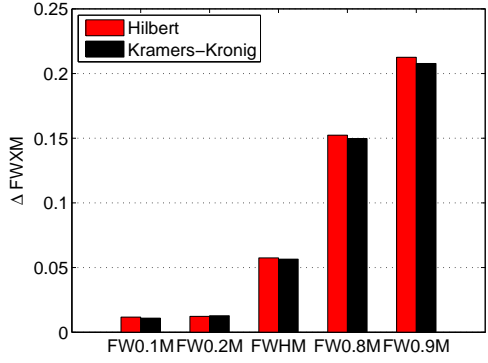} 
  \caption{$\Delta_{FWXM}$  for 1000 profiles with both methods.}%VH add  picture 
   \label{fwxm}
\end{figure}

\begin{figure}[htbp]
 \centering
  \includegraphics*[width=65mm]{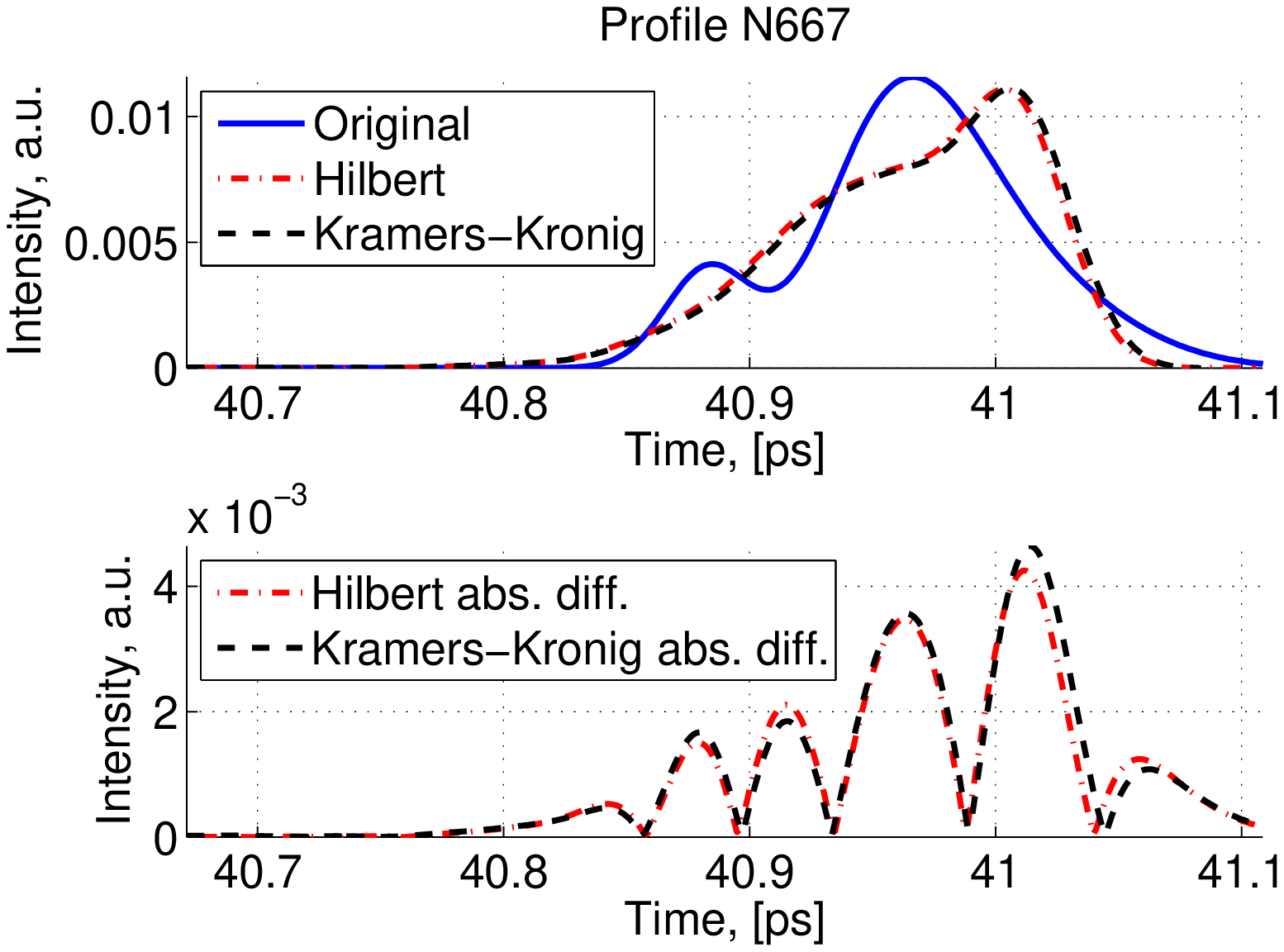}\\
    \includegraphics*[width=65mm]{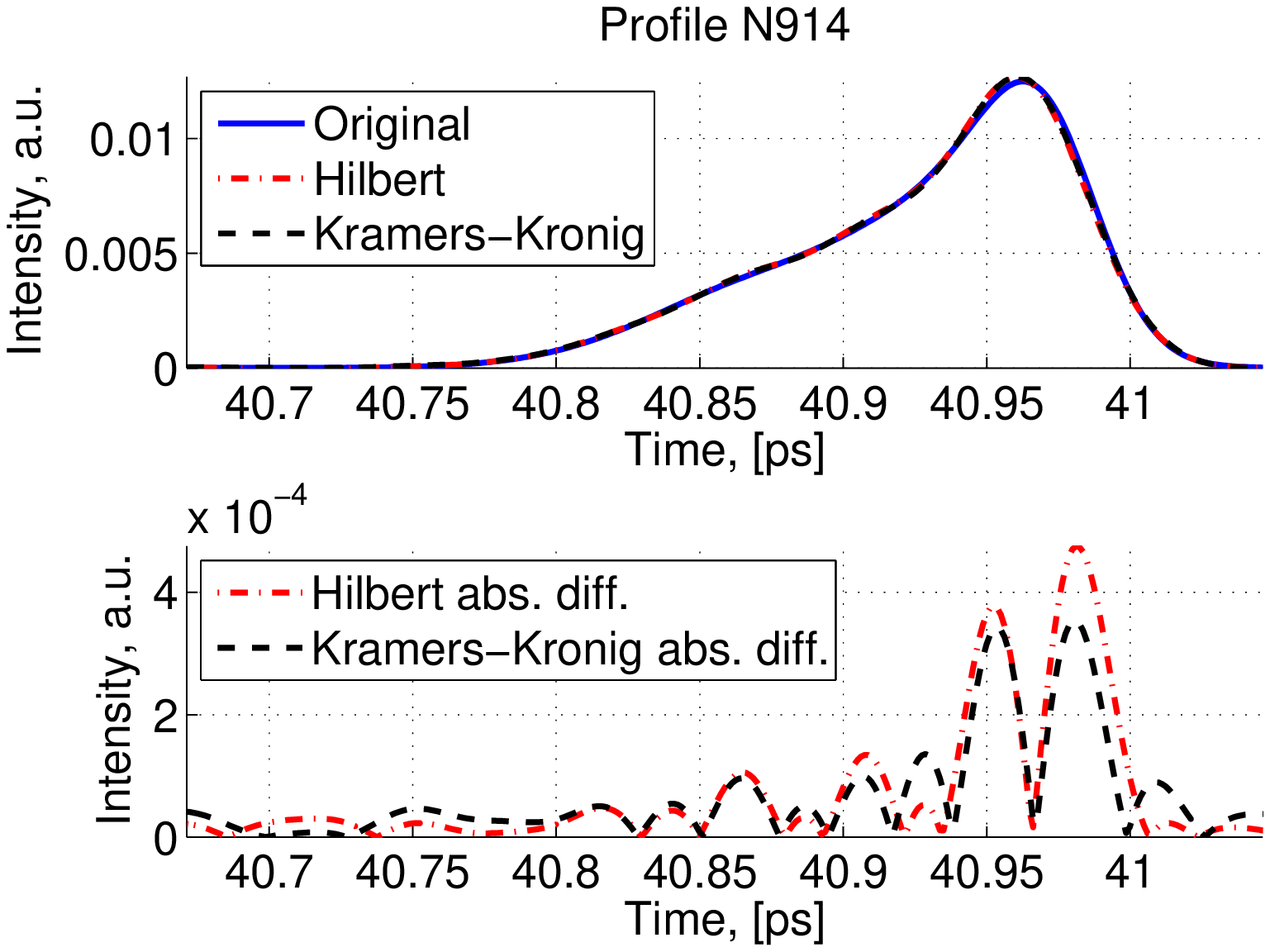}
    \caption{Original and reconstructed profile and their difference for bad profile (top) and good profile (bottom).}
   \label{mod}
\end{figure}

The fact that the phase recovery method based on the Kramers-Kronig relation gives a worst $\chi^2$  than the method based on Hilbert relation has been investigated. It is caused by the presence of negative components in the tails of the profiles. Figure~\ref{expKK} highlights this issue for one of the profiles.
% VH add this block
%%%%%%%%%%%%%%%%%%%%%%%%%%%%%%%%%
Figure~\ref{fwxm} shows the different FWXH values   for different values of X. This shows that at   different height of the profiles the quality of reconstruction varies: there is a better agreement in the tails (X=10\%) than at the top of the profile (X=90\%).
 Figure~\ref{mod} shows the modulus of the difference between the original and reconstructed profiles. One can see oscillations in the difference between the original and reconstructed profile.

\begin{figure}[htbp]
 \centering
  \includegraphics*[width=70mm]{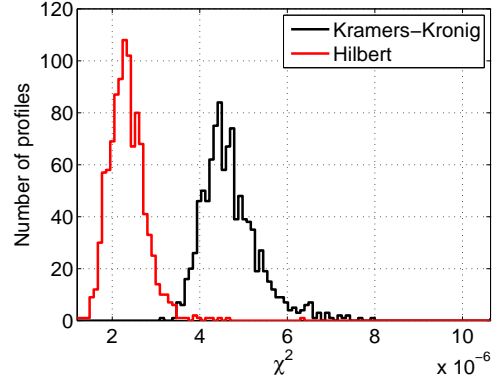} \\
  \includegraphics*[width=70mm]{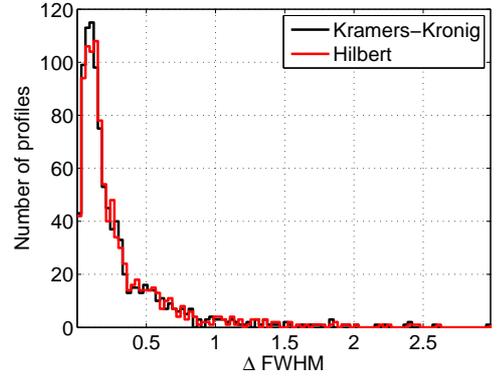} \\
  \caption{Distribution of the $\chi^2$ in the case of a Lorenzian distribution. }
   \label{lorenz}
\end{figure}

While doing this work we also became aware of the discussion in~\cite{Pelliccia:2014vba} where it is argued that these reconstruction method have more difficulties with lorentzian profiles  than gaussian profiles. Therefore we simulated 1000 Lorenzian profiles and performed a similar study. This is shown in figure~\ref{lorenz}. Although the $\chi^2$ is slightly worse in that case than in the case of gaussian profiles there still a good agreement between the original and reconstructed profiles.

\begin{figure}[htbp]
 \centering
  \includegraphics*[width=70mm]{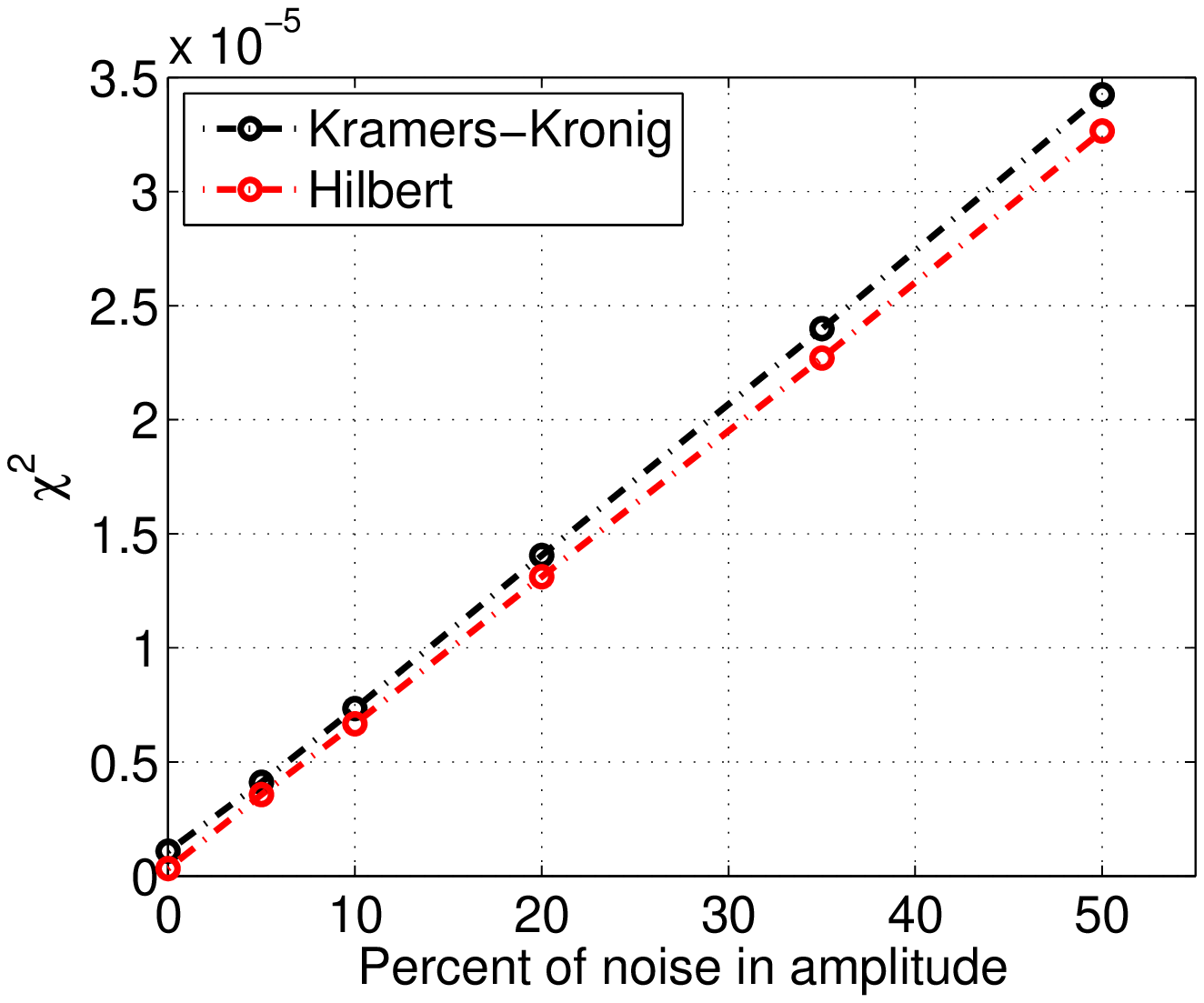}\\
  \includegraphics*[width=70mm]{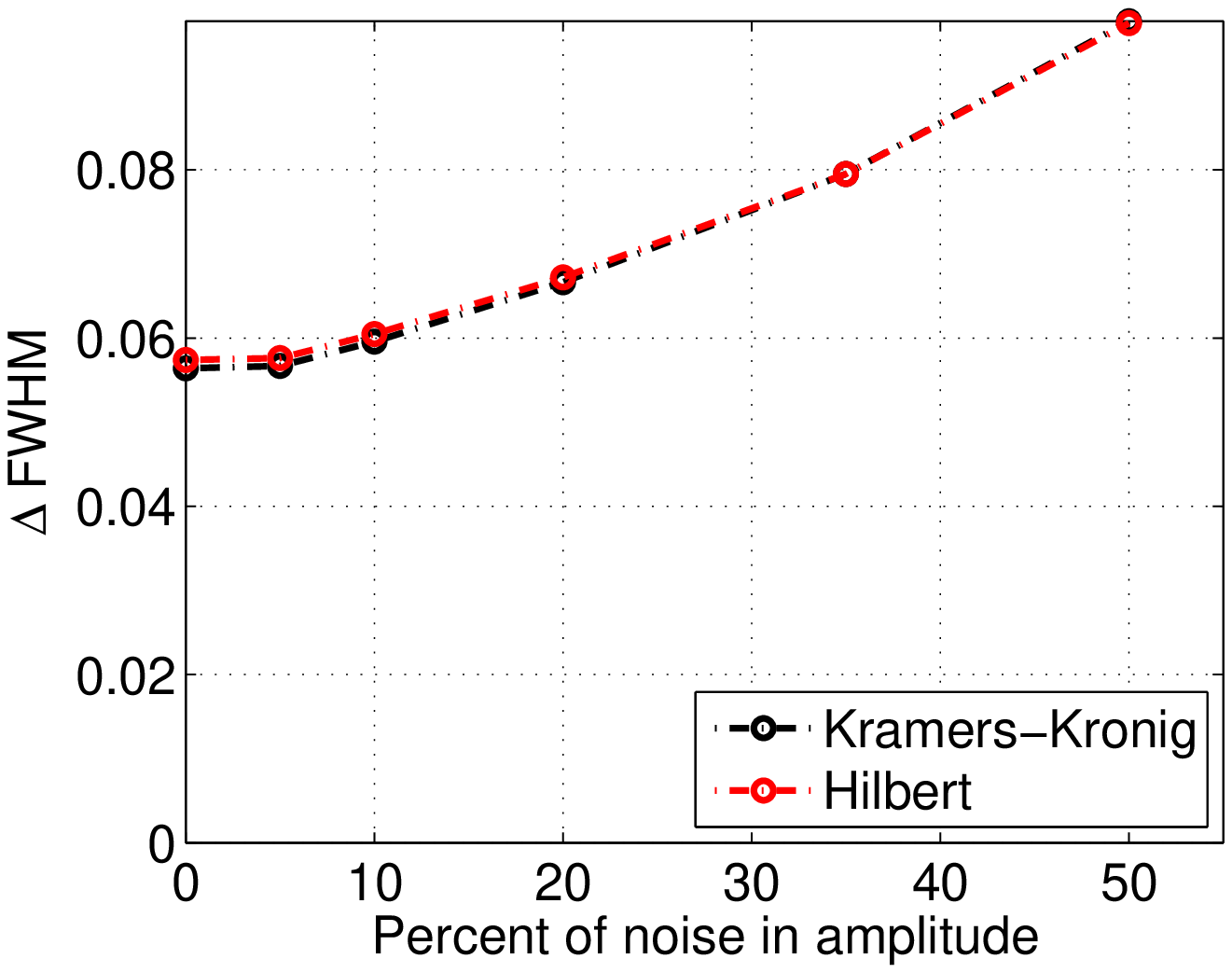}\\
  \caption{Mean $\chi^2$ and $\Delta_{FWXM}$  as function of noise amplitude.}
   \label{noise}
\end{figure}

% Effect of noise.
In our discussion so far we considered only the ideal case where no noise is added to the measured spectrum. However in a real experiment a noise component has to be added to the measured spectrum. This noise was added as follow :
\begin{equation}
O_i' = O_i \times [1 + ( n_i N_{max}) ] 
\end{equation}
 where $O_i$ is the observed value,  $O_i'$ is the observed value with noise, $n_i$ is a random number between 0 and 1 (all numbers between 0 and 1 being equiprobable), and $N_{max}$ is the 
maximum noise for that simulation (depending on the case this can be 5\%, 10\%, 20\%, 30\%, 40\% or  50\%). This study was done using linear sampling with 33 samples  and 1000 simulated profiles for each noise value. Figure~\ref{noise} shows how the $\chi^2$ is modified when this noise component is added.
 
\section{Discussion}

We performed extensive simulation to estimate the performance of two phase recovery methods in the case of multi-gaussian and Lorenzian profiles. In both cases we found that when the sampling frequencies are chosen correctly  we obtained a good agreement between the original and reconstructed profiles (in most cases $\Delta_{FWXM} < 10\%$;  $\chi^2 \sim 10^{-6}$). This confirms that such methods are suitable to reconstruct the longitudinal profiles measured at particle accelerators using radiative methods.

The authors are grateful for the funding received from the French ANR (contract ANR-12-JS05-0003-01), the PICS (CNRS) "Development of the instrumentation for accelerator experiments, beam monitoring and other applications" and  Research Grant \#F58/380-2013 (project F58/04) from the State Fund for Fundamental Researches of Ukraine in the frame of the State key laboratory of high energy physics and the IDEATE International Associated Laboratory (LIA).

%\bibliographystyle{unsrt}
%\bibliography{biblio}

%%\begin{thebibliography}{9}   % Use for  1-9  references
%\begin{thebibliography}{99} % Use for 10-99 references

%\bibitem{accelconf-ref}
%	C. Petit-Jean-Genaz and J. Poole,
%	``JACoW, A service to the Accelerator Community,''
%	EPAC'04, Lucerne, July 2004, THZCH03,  p.~249,
%	\url{http://www.JACoW.org/e04/papers/THZCH03.PDF}

% \end{thebibliography}

\end{document}